\documentclass[sigconf]{acmart}
\usepackage{graphicx} 
\usepackage{mathtools}
\usepackage{multirow}
\usepackage{array}
\usepackage{booktabs}
\usepackage{caption}
\usepackage{tabularx}
\usepackage{enumitem}
\usepackage{listings}
\usepackage{xcolor}
\usepackage{tikz}
\usepackage{graphicx}
\usepackage{etoolbox}
\usepackage{epigraph}
\setcopyright{rightsretained}
\usepackage[capitalize,noabbrev]{cleveref}
\definecolor{orange}{rgb}{0, 0, 0}   
\colorlet{ORANGE}{orange}

\definecolor{OCircle}{HTML}{AB73FF}
\definecolor{CCircle}{HTML}{3CC5F2}
\definecolor{GlobalCircle}{HTML}{4c78a8}
\definecolor{DragCircle}{HTML}{54a24b}
\definecolor{SemanticCircle}{HTML}{b279a2}
\definecolor{RegenerateCircle}{HTML}{eeca3b}
\definecolor{EditCircle}{HTML}{bab0ac}
\definecolor{BlueCircle}{HTML}{3366D4}

 \newcommand{\globalcircle}{%
\begin{tikzpicture}[enum/.style={circle,draw=gray,very
    thin,fill=GlobalCircle,text=white,inner sep=1pt,minimum size=9pt},baseline=-3pt]
  \node (node 1) at (0,0) [enum] {};%
 \end{tikzpicture}}

\newcommand{\dragcircle}{%
\begin{tikzpicture}[enum/.style={circle,draw=gray,very
    thin,fill=DragCircle,text=white,inner sep=1pt},minimum size=9pt,baseline=-3pt]
  \node (node 1) at (0,0) [enum] {};%
 \end{tikzpicture}}
\newcommand{\semanticcircle}{%
\begin{tikzpicture}[enum/.style={circle,draw=gray,very
    thin,fill=SemanticCircle,text=white,inner sep=1pt},minimum size=9pt,baseline=-3pt]
  \node (node 1) at (0,0) [enum] {};%
 \end{tikzpicture}}
\newcommand{\regeneratecircle}{%
\begin{tikzpicture}[enum/.style={circle,draw=gray,very
    thin,fill=RegenerateCircle,text=white,inner sep=1pt},minimum size=9pt,baseline=-3pt]
  \node (node 1) at (0,0) [enum] {};%
 \end{tikzpicture}}
\newcommand{\editcircle}{%
\begin{tikzpicture}[enum/.style={circle,draw=gray,very
    thin,fill=EditCircle,text=white,inner sep=1pt},minimum size=9pt,baseline=-3pt]
  \node (node 1) at (0,0) [enum] {};%
 \end{tikzpicture}}

\newcommand{\ccircle}[1]{%
\begin{tikzpicture}[enum/.style={circle,fill=CCircle,text=white,inner sep=1pt},baseline=-3pt]
  \node (node 1) at (0,0) [enum] {\scriptsize #1};%
\end{tikzpicture}}

\AtBeginDocument{%
  \providecommand\BibTeX{{%
    \normalfont B\kern-0.5em{\scshape i\kern-0.25em b}\kern-0.8em\TeX}}}

\copyrightyear{2025}
\acmYear{2025}

\usepackage{etoolbox}
\makeatletter
\tracingpatches
\patchcmd{\maketitle}{\@copyrightpermission}{
   \begin{minipage}{0.3\columnwidth}
     \href{<https://creativecommons.org/licenses/by/4.0/>}{\includegraphics[width=0.90\textwidth]{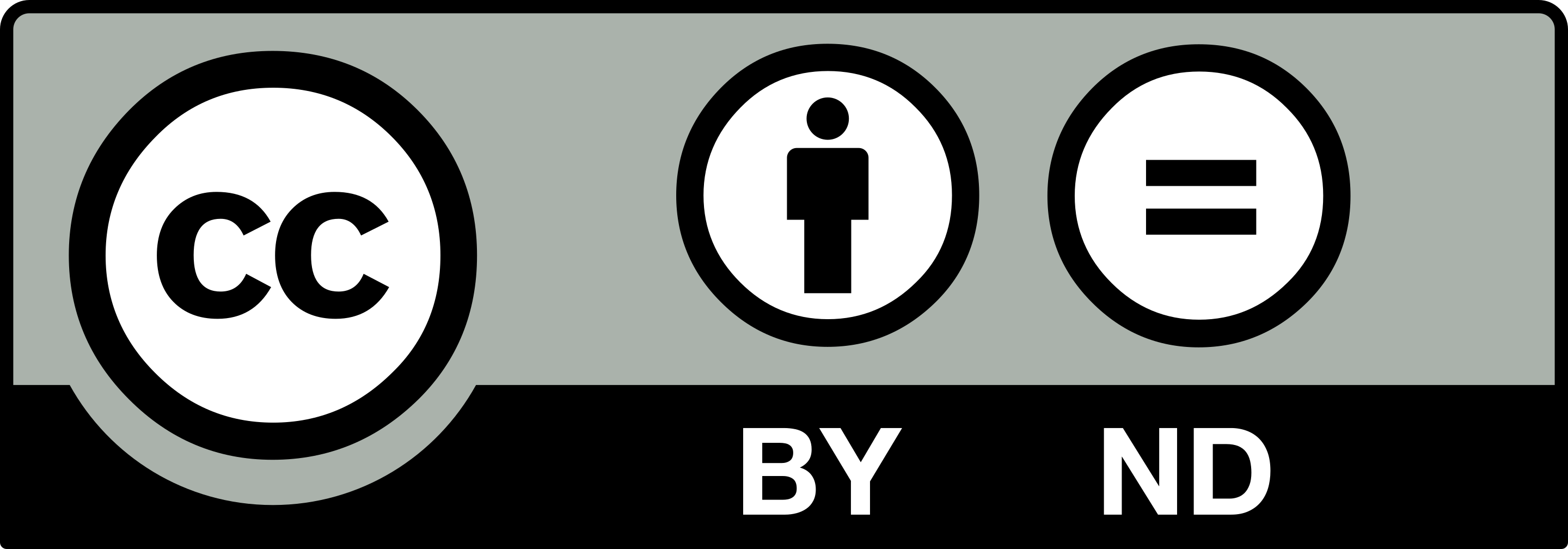}}
   \end{minipage}\hfill
   \begin{minipage}{0.7\columnwidth}
     \href{<https://creativecommons.org/licenses/by-nd/4.0/>}{This work is licensed under a Creative Commons Attribution-NoDerivs International 4.0.}
   \end{minipage}

   \vspace{5pt}
}{}{}

\makeatletter

\acmConference[CHI '25]{CHI Conference on Human Factors in Computing Systems}{April 26-May 1, 2025}{Yokohama, Japan}
\acmBooktitle{CHI Conference on Human Factors in Computing Systems (CHI '25), April 26-May 1, 2025, Yokohama, Japan}\acmDOI{10.1145/3706598.3713924}
\acmISBN{979-8-4007-1394-1/25/04}

\newcommand{\sys}[0]{\textsc{Misty}}

\newcommand{\sourceNode}{source node}
\newcommand{\exampleNode}{example node}

\begin{document}

\title[]{\sys{}: UI Prototyping Through Interactive Conceptual Blending}

\author{Yuwen Lu}
\authornote{Work done during internship at Apple.}
\affiliation{%
  \institution{University of Notre Dame}
  \city{Notre Dame}
  \state{IN}
  \country{USA}}
\email{ylu23@nd.edu}

\author{Alan Leung}
\affiliation{%
  \institution{Apple}
  \city{Seattle}
  \state{WA}
  \country{USA}}
\email{alleu@apple.com}

\author{Amanda Swearngin}
\affiliation{%
  \institution{Apple}
  \city{Seattle}
  \state{WA}
  \country{USA}}
\email{aswearngin@apple.com}

\author{Jeffrey Nichols}
\affiliation{%
  \institution{Apple}
  \city{Seattle}
  \state{WA}
  \country{USA}}
\email{jwnichols@apple.com}

\author{Titus Barik}
\affiliation{%
  \institution{Apple}
  \city{Seattle}
  \state{WA}
  \country{USA}}
\email{tbarik@apple.com}

\begin{teaserfigure}\includegraphics[trim=0cm 0cm 0cm 0cm, clip=true, width=\textwidth]{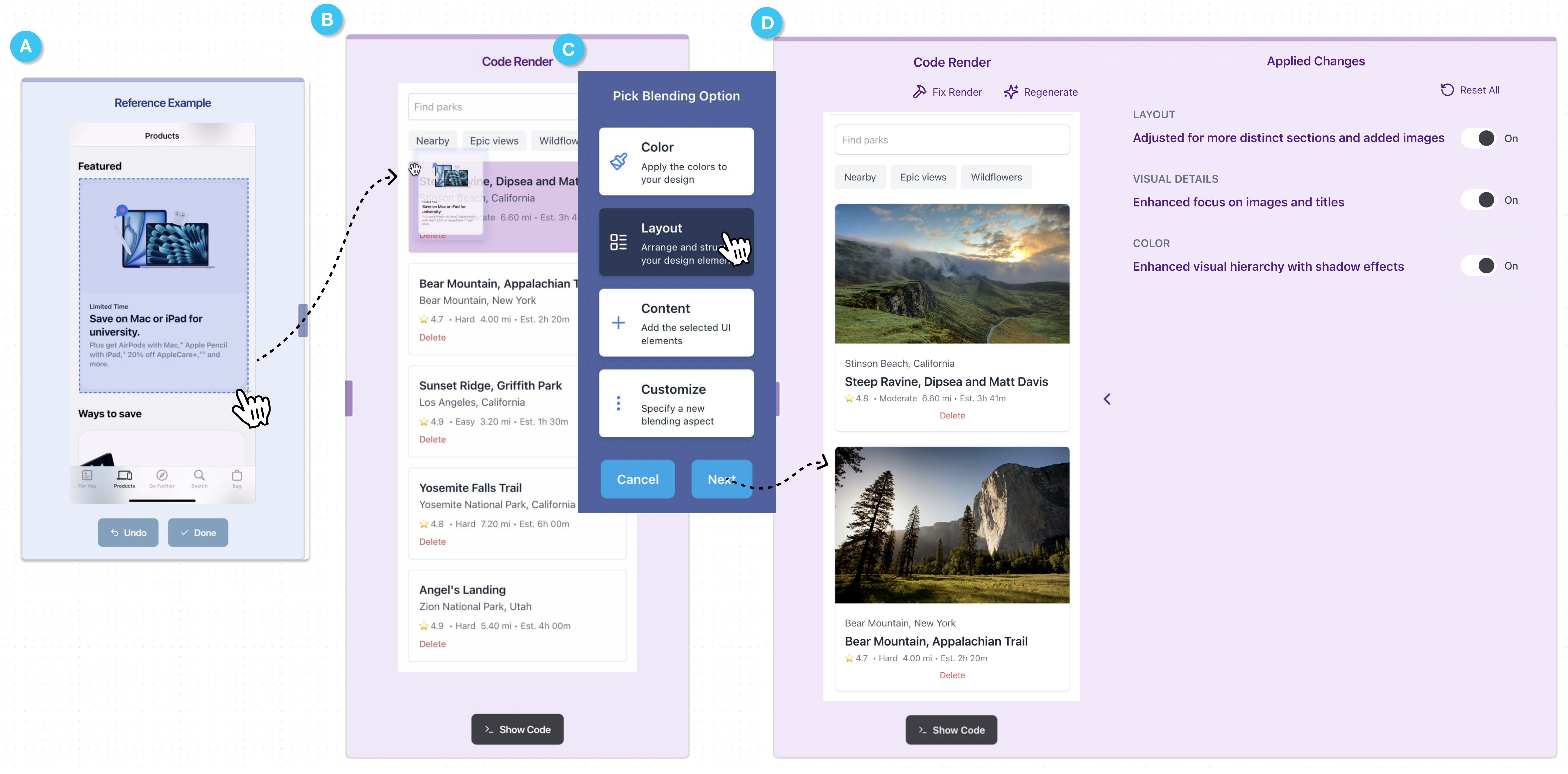}
    \caption{The \sys{} system. \textmd{\sys{} embodies the interactive conceptual blending interaction for UI prototyping, where UI developers \emph{blend} selected sections of an example image \protect\ccircle{A} into work-in-progress UI code \protect\ccircle{B}. Users can \emph{specify} aspect(s) of the image to blend in \protect\ccircle{C}. After viewing the blending output \protect\ccircle{D}, the user can further tweak details by toggling on/off within the \emph{semantic diff} (that is, semantically categorized code changes) between the updated and original UI versions. Initially, the work-in-progress UI shows basic cards with some park information. The user selects the card layout from the example UI, drags it onto their current design's cards, and chooses to blend the "Layout". \sys{} then generates updated UI code, preserving the original content but presenting it in a new card layout with images and hierarchical text descriptions, mirroring the example's layout.}}
    \Description{This figure illustrates the \sys{} system, an interactive conceptual blending tool for UI prototyping. The image is divided into four main sections labeled A through D. Section A shows a reference example UI with a featured product and navigation options. Section B displays a code render of a work-in-progress UI for finding parks. Section C presents a ``Pick Blending Option'' menu with choices for Color, Layout, Content, and Customize. Section D shows the blending output with an updated UI design and a list of applied changes. The accompanying text explains that \sys{} allows UI developers to blend selected sections from an example image (A) into their work-in-progress UI code (B). Users can specify which aspects of the image to blend in the menu (C). After viewing the blending output (D), users can fine-tune details by toggling on/off within the semantic diff, which shows semantically categorized code changes between the updated and original UI versions. This process demonstrates \sys{}'s capability to facilitate rapid, interactive UI prototyping through conceptual blending of design elements.}
    \label{fig:misty-teaser}
\end{teaserfigure}

\begin{abstract}
UI prototyping often involves iterating and blending elements from examples such as screenshots and sketches, but current tools offer limited support for incorporating these examples. Inspired by the cognitive process of conceptual blending, we introduce a novel UI workflow that allows developers to rapidly incorporate diverse aspects from design examples into work-in-progress UIs. We prototyped this workflow as \sys{}. Through a exploratory first-use study with 14 frontend developers, we assessed \sys{}'s effectiveness and gathered feedback on this workflow. Our findings suggest that \sys{}'s conceptual blending workflow helps developers kickstart creative explorations, flexibly specify intent in different stages of prototyping, and inspires developers through serendipitous UI blends. \sys{} demonstrates the potential for tools that blur the boundaries between developers and designers.
\end{abstract}



\settopmatter{printfolios=true}

\maketitle

\section{Introduction}

\renewcommand{\epigraphflush}{center}
\begin{epigraphs}
\qitem{
I can't speak, afraid to jinx it\\
I don't even dare to wish it\\
But your eyes are flying saucers from another planet\\
Now I'm all for you like Janet\\
Can this be a real thing? Can it?
}{Taylor Swift, \emph{Snow on the Beach} (2023)}
\end{epigraphs}

When prototyping user interfaces (UIs), developers routinely turn to reference examples---such as sketches, screenshots, or existing design languages---as sources of inspiration. These examples aid developers in ideation~\cite{herring2009getting, bunian2021vins, chen2019gallery, lu2024flowy}, help them to avoid design fixation~\cite{dow2010parallel}, and spur creative exploration~\cite{o2015designscape, ritchie2011d}. As with other forms of creative expression, developers then selectively iterate and blend various elements from these examples---such as layout or color---in this case, to design and produce a novel, distinctive UI.

This imaginative cognitive process of \emph{conceptual blending} plays a fundamental role in the construction of meaning in everyday life, and it is so natural to the way we reason that we are often not even consciously aware that we are blending. ``The essence of the operation,'' as described by Fauconnier and Turner, ``is to construct a partial match between two input mental spaces, to project selectively from those inputs into a novel `blended' mental space, which then dynamically develops emergent structure''~\cite{fauconnier2003conceptual}. For example, when singer-and-songwriter Taylor Swift tells us ``your eyes are flying saucers from another planet,'' we intuit that the eyes are not literally flying saucers, but instead readily project selectively the two inputs---eyes and flying saucers---to blend that their gaze has a mesmerizing and otherworldly quality~\cite{santika2023analysis}.

While it's unsurprising that other research has demonstrated the utility of conceptual blending in computational creativity support for various domains~\cite{camara2007creativity, zbikowski2018conceptual}, there are nevertheless limited tools for blending user interfaces. \textcolor{orange}{Blending visual inspirations is common in UI creation, but designers and developers often have to manually implement these blends from scratch~\cite{herring2009getting}.} How then, can we bring conceptual blending into our tools for user interface creation?

Towards addressing this gap, we created a web prototype system, \sys{}\footnote{\sys{} is an acronym for Mixing Interfaces by Semantic Transformation. Plus Y, because why not.}, that embodies and operationalizes conceptual blending. With \sys{}, users can upload example UI screenshots or hand-drawn sketches, specify specific regions and aspects of interest, blend them into their work-in-progress UIs, and use a \emph{semantic diff} to make additional refinements to the blend  (\cref{fig:misty-teaser}). We postulated that because generative AI models encode certain characteristics of human ingenuity\textcolor{orange}{~\cite{lakoff2014mapping, tong2024metaphorunderstandingchallengedataset}}, these latest multi-modal AI models would have within them some capacity to automatically interpret and apply conceptual blending. \textcolor{orange}{In \sys{}, we evaluated conceptual blending in developing individual mobile UI screens.}

To evaluate \sys{}, we conducted \textcolor{orange}{a user study} with 14 frontend developers. The study results revealed three key insights: First, participants found \sys{}'s varied levels of input granularity valuable, supporting different stages of the design process from early ideation to detailed refinement. Second, while the current AI models showed limitations in generating perfect conceptual blends, participants appreciated our semantic diff and the code editor to tweak generation results, often using imperfections as springboards for further ideation. Third and finally, conceptual blending through \sys{} enhanced UI prototyping workflows by supporting both focused iteration on single designs and broad exploration of diverse alternatives.\looseness=-1

The contributions of this paper are:
\begin{itemize}
    \item A novel UI prototyping approach, conceptual blending, which integrates diverse, meaningful aspects of reference examples into the user's current UI.
    \item \sys{}, a system that supports conceptual blending, where UI developers interactively blend example images into work-in-progress code.
    \item User studies that demonstrated the value of \sys{} to kickstart and inspire UI development, and dissolve designer---developer boundaries.
\end{itemize}

\section{Example Usage Scenario For \sys{}}

Fuji is a frontend developer creating a user profile management page in a new mobile app. Using React Native and Tailwind CSS, Fuji has laid out some basic components for the profile page, including a circular avatar, the user's name, a brief bio, contact information, and a list of the user's interests. While functional, Fuji feels the current UI is too basic and lacks visual appeal. The layout is simple, the color scheme is monotonous, and there's no clear visual hierarchy. Fuji decides to look online for inspiration to enhance the design.

Fuji has recently installed \sys{}, which streamlines the incorporation of design examples through conceptual blending. \textcolor{orange}{With some inspiring UI designs found online, Fuji is ready to use \sys{} to explore improvement directions for the profile UI.}

\begin{center}
\includegraphics[width=\linewidth]{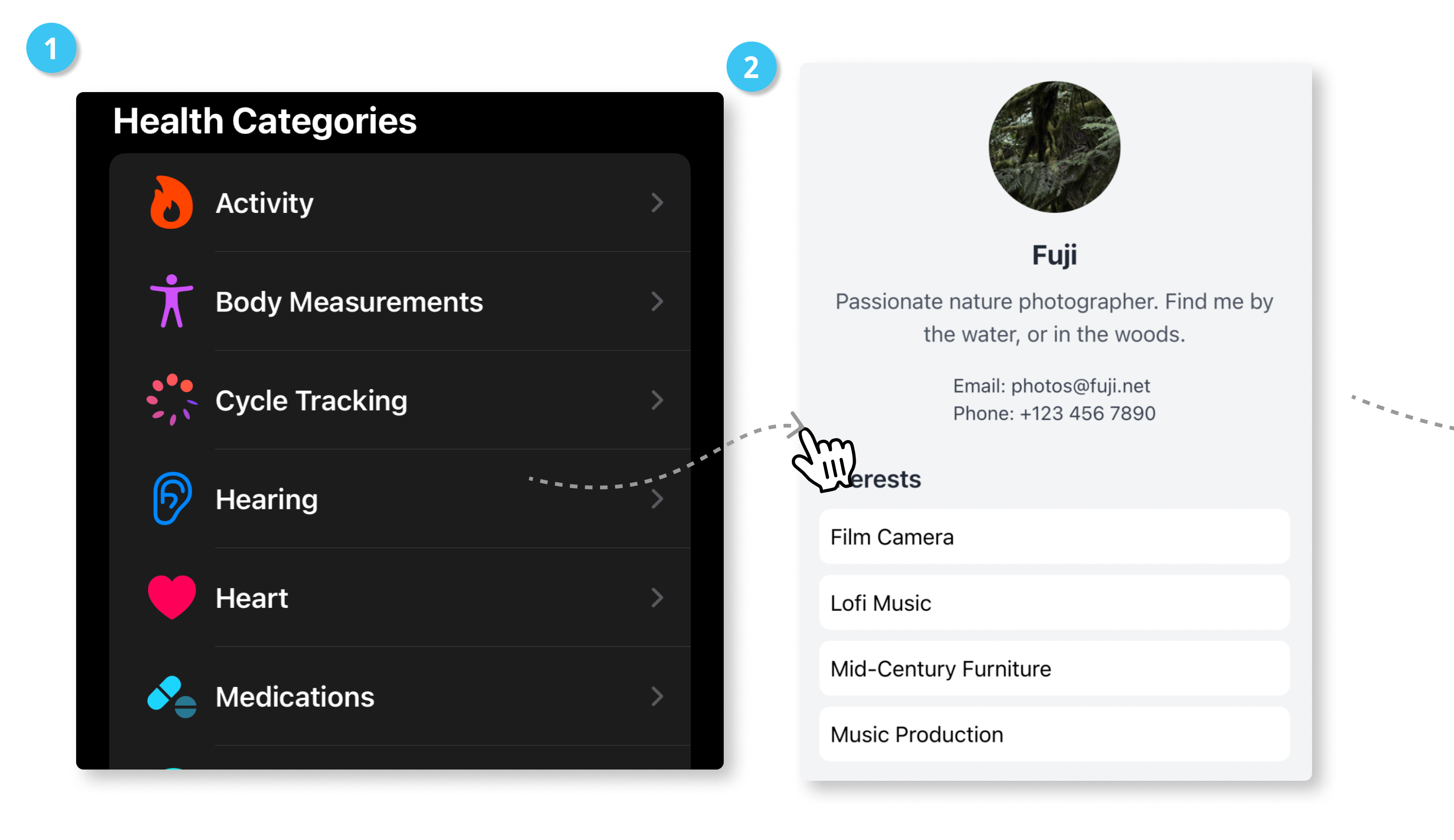}
\Description{This figure demonstrates \sys{}'s global blending feature for initial design exploration. The image is divided into two sections: section 1 shows an example UI with ``Health Categories'' in a dark-themed list, while section 2 displays Fuji's current profile UI with a light theme. A dotted line with a hand cursor connects these two sections, illustrating the blending process. The accompanying text explains that Fuji began by uploading their current profile UI code and screenshots of inspiring designs to \sys{}. To explore possibilities quickly, Fuji utilized \sys{}'s global blending feature, connecting the example UI (1) to their source UI (2). This action initiated a global, screen-level blending operation. \sys{} then generated updated code that incorporated Fuji's existing profile content while adopting the visual styles of the example UI. This process demonstrates how \sys{} enables users to efficiently experiment with different design aesthetics while maintaining their original content, facilitating rapid UI prototyping and exploration.}
\end{center}

\paragraph{Global blending for initial exploration}
Fuji began by uploading the code for their current profile UI to \sys{}, along with screenshots of a few inspiring profile UI designs. \textcolor{orange}{To get an overview of possibilities, Fuji used \sys{}'s global blending feature to quickly explore alternatives.} They connected the example UI \protect\ccircle{1} to their source UI \protect\ccircle{2}, triggering a global, screen-level blending operation. \sys{} generated updated code that took Fuji's current profile UI content and \textit{blended} into it the visual styles of the example UI  \protect\ccircle{3}.

\begin{center}
\includegraphics[width=\linewidth]{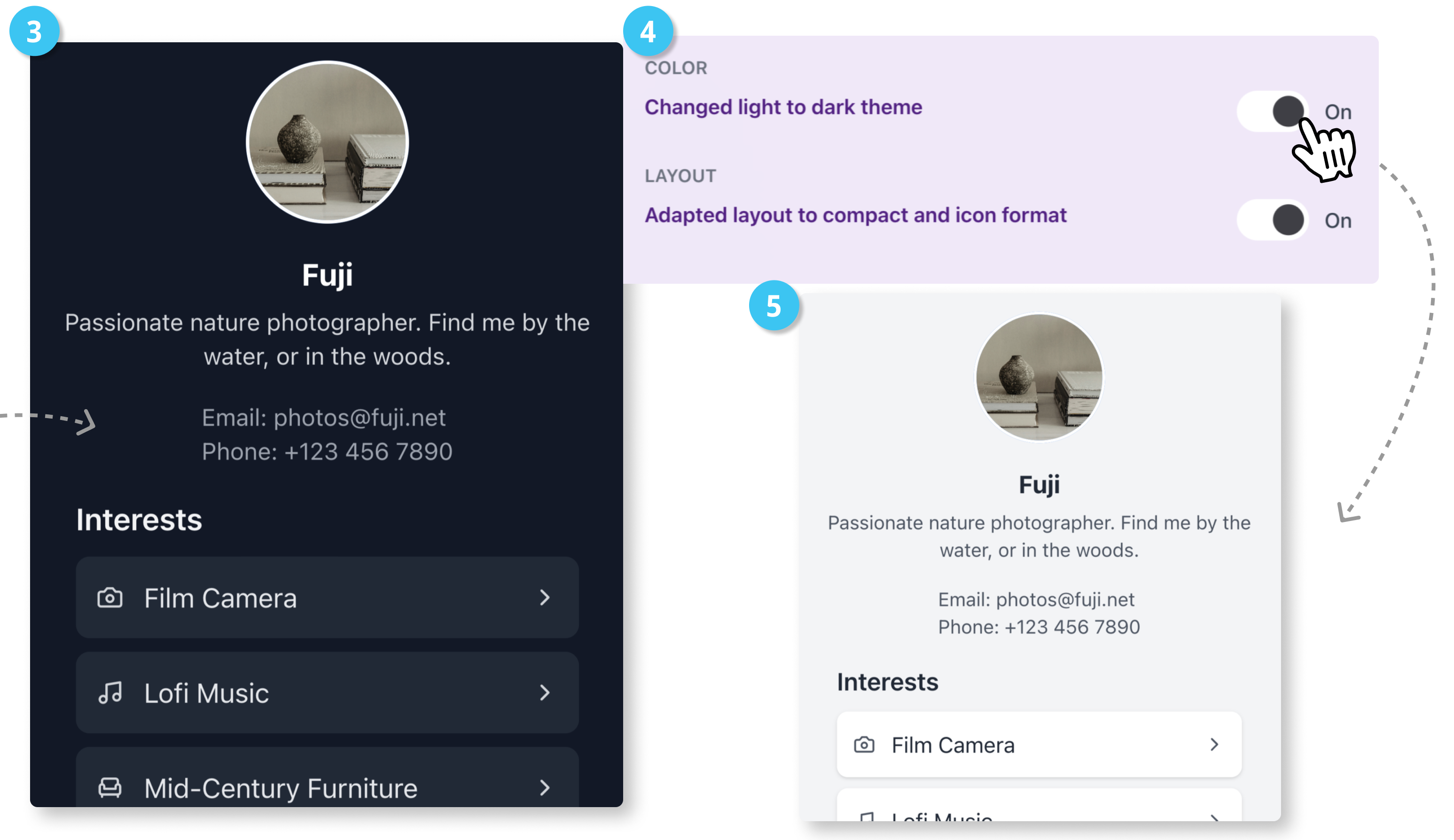}
\Description{This figure illustrates the iterative design process facilitated by \sys{}. The image is divided into three sections labeled 3, 4, and 5. Section 3 shows Fuji's profile UI with a dark theme and compact list layout for interests. Section 4 displays a summary of changes made by \sys{}, including the switch to a dark theme and adaptation of layout to a compact, icon format. Two toggle switches are shown, with the cursor hovering over the first one. Section 5 presents the final iteration of Fuji's profile, reverting to a light theme while maintaining the compact list layout for interests. The accompanying text explains that \sys{} provided a summary of code updates alongside the outputs. Fuji observed that \sys{} had applied a dark theme and adopted the example's list item layout. Preferring to keep the updated layout but revert to a light theme, Fuji toggled off the dark theme option. In response, \sys{} generated an updated UI that preserved the new layout while returning to the original light theme}
\end{center}

\sys{} provided a list of changes alongside the outputs \protect\ccircle{4}, summarizing the code updates. Fuji noticed that \sys{} helped to turn their UI into dark theme and adopt the list item layout in the example. Fuji decided to keep the updated list item layout, but turn off the dark theme. By switching off the toggle corresponding to dark theme, Fuji received an updated result from \sys{} \protect\ccircle{5}, which returned the UI back to the previous light theme but kept the other changes. \textcolor{orange}{This gave Fuji more control in exploration, by picking the desired categories of change and directly seeing the visual result.} \looseness=-1

Usually, Fuji would manually re-create the example using code, which is very time-consuming. Moreover, without seeing their own content reflected in these design variations, Fuji would struggle to get a direct impression of how each design would work in their scenario. But now, with \sys{}, Fuji can get a quick visual preview of multiple design possibilities, all populated with their own content. \textcolor{orange}{\sys{}'s global blending feature supports a more comprehensive exploration of the design space.}

\begin{center}
\includegraphics[width=\linewidth]{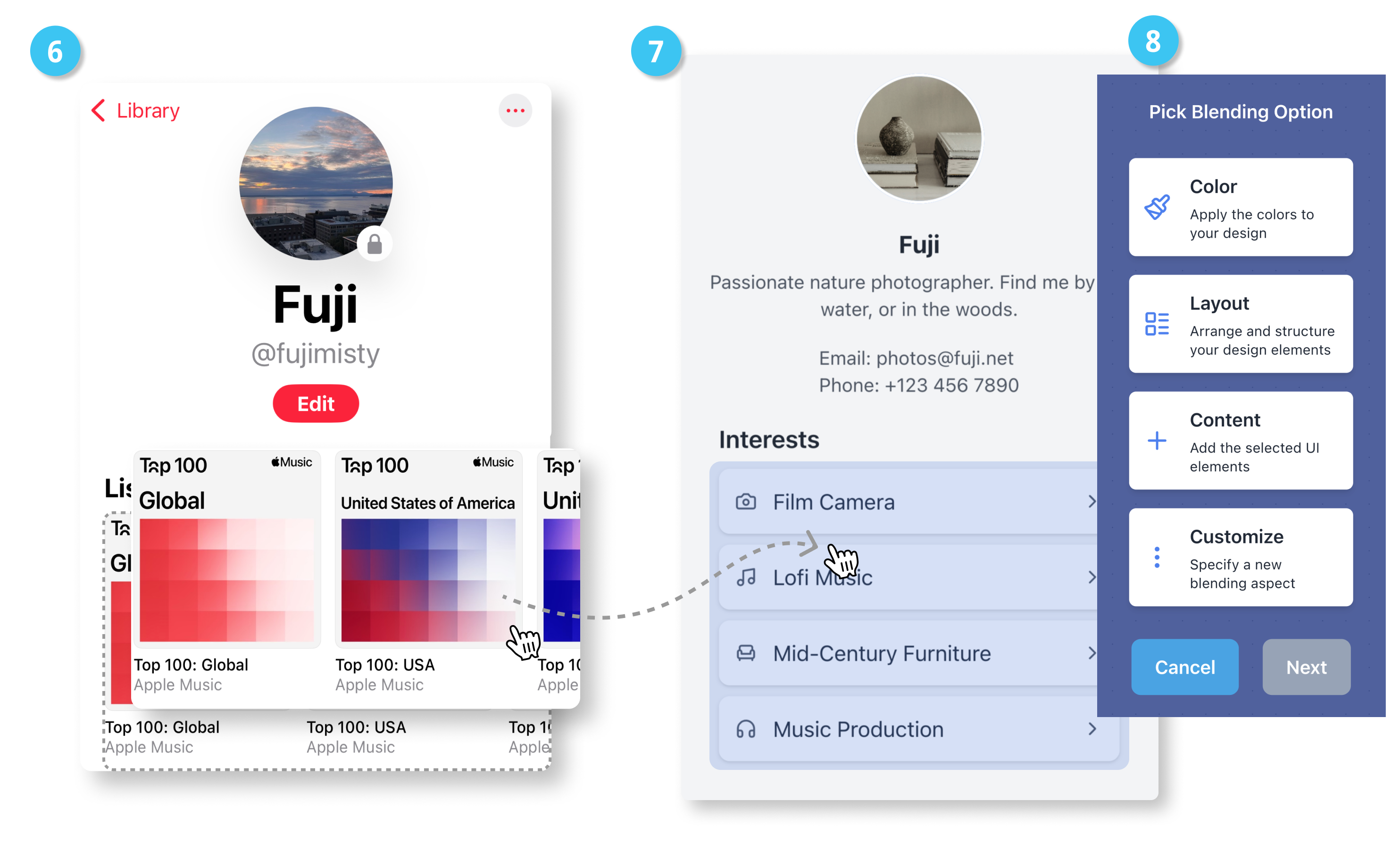}
\Description{This figure illustrates the user interface for \sys{}, focusing on a profile customization process. The image is divided into three main sections, numbered 6, 7, and 8. Section 6 shows a profile editing screen with a circular profile picture of a sunset over water, the name ``Fuji'' and username ``fujimist''. Below this is a horizontally scrollable list of music playlists labeled ``Top 100'' for different regions. Section 7 displays a profile card for Fuji, describing them as a ``Passionate nature photographer''. It includes contact information and a list of interests such as ``Film Camera'', ``Lofi Music'', ``Mid-Century Furniture'', and ``Music Production''. Section 8 presents a ``Pick Blending Option''menu with four choices: Color, Layout, Content, and Customize. Each option has a brief description of its function. The image demonstrates the process of dragging elements from one design (the music playlists) to another (the interests list), as indicated by a dotted line with a hand cursor icon. This visualization corresponds to the paragraph in the second image, which explains how Fuji, the user, is using the application's drag-and-drop feature to incorporate a horizontal scrollable carousel layout from an example UI into the ``Interest'' section of their own profile design.}
\end{center}

\paragraph{Targeted blending for sections of interest}
As Fuji looked through other examples, they decided to adopt a design that is good at presenting information in a clean, card-based layout using a horizontal scrollable carousel \protect\ccircle{6}. Fuji decided to incorporate this layout more precisely into the ``Interest'' section of their UI. Using \sys{}'s drag-and-drop feature, Fuji selected the horizontal scrollable list from the example UI \protect\ccircle{6}, then dragged it onto the corresponding area in their current design \protect\ccircle{7}.
\begin{center}
\includegraphics[width=\linewidth]{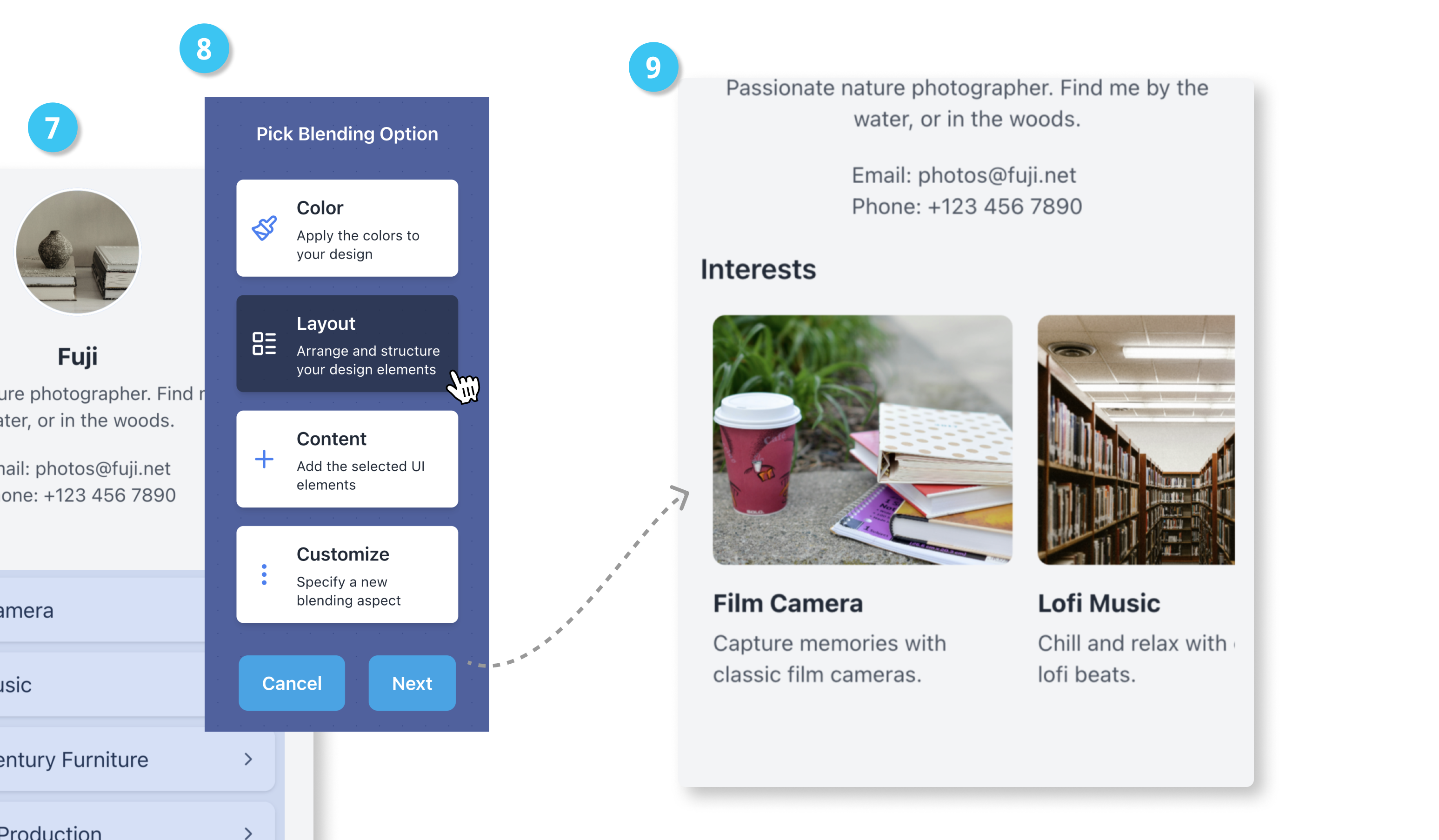}
\Description{This figure illustrates the interface of \sys{} during a profile customization process. The image is divided into three sections numbered 7, 8, and 9. Section 7 shows part of Fuji's profile with their photo and bio. Section 8 displays a ``Pick Blending Option'' menu where the ``Layout'' option is highlighted, indicating Fuji's choice to blend the layout of the example UI. Section 9 showcases the result of this blending operation: Fuji's updated ``Interests'' section now features a horizontally scrollable list with images and descriptions for each interest, such as ``Film Camera'' and ``Lofi Music''. A dotted arrow connects the layout option to the new interests display, visualizing the transformation process. The accompanying text explains that after Fuji selected the layout blend option, \sys{} generated this new version of the interests section, incorporating a sleek, horizontally scrollable list that includes contextually relevant details for each interest item, effectively adapting the example UI's layout to Fuji's profile content.}
\end{center}

 In the popup that appeared \protect\ccircle{8}, Fuji specified that they wanted to blend the ``layout'' of the example UI. \sys{} processed this request and generated a new version of the interests section \protect\ccircle{9}, featuring a sleek, horizontally scrollable list of users' interest items, while also adding contextually relevant details of each interest to fit in the example UI.

\begin{center}
\includegraphics[width=\linewidth]{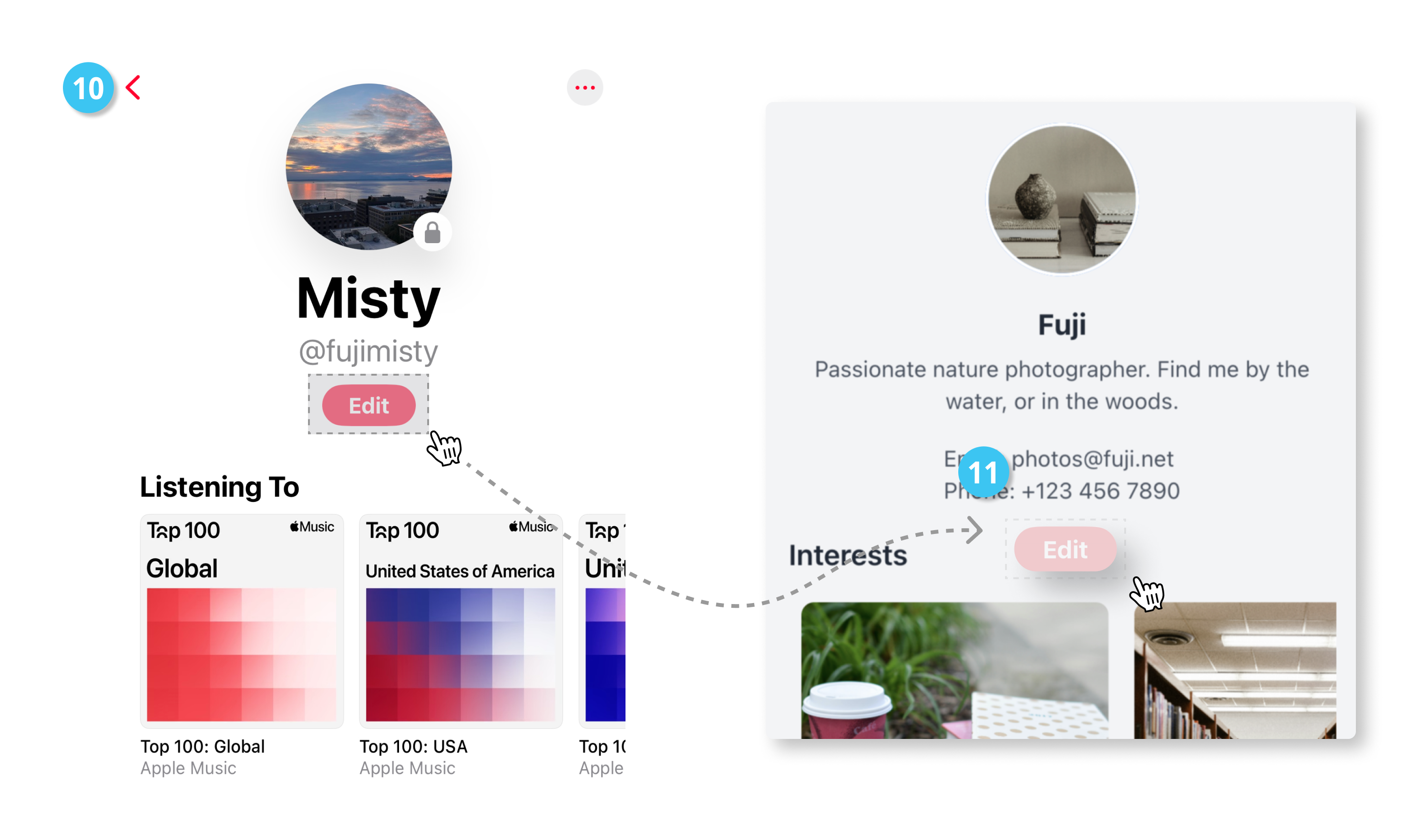}
\Description{This figure demonstrates how \sys{} facilitates the addition of new UI elements through blending. The image is split into two sections, labeled 10 and 11. Section 10 shows an example UI profile for ``Misty'' with an ``Edit'' button beneath the username. Section 11 displays Fuji's profile, now featuring a newly added ``Edit'' button next to the ``Interests'' section. A dotted arrow connects the ``Edit'' buttons in both sections, illustrating the drag-and-drop process. The accompanying text explains that while refining their design, Fuji decided to incorporate the ``Edit'' button from the example UI (section 10) into their own profile design (section 11). This addition wasn't part of Fuji's original layout, but they recognized its potential value for allowing users to update their profiles. Using \sys{}'s drag-and-drop functionality, Fuji transferred this UI element from the example to the appropriate area in their current design, demonstrating how \sys{} enables users to easily augment their designs with new, useful features through intuitive blending operations.}
\end{center}

\paragraph{Adding new UI elements through blending}

While refining their design, Fuji realized it would be beneficial to add a new ``edit'' button they saw in the previous example UI \protect\ccircle{10}, to let users update their profile. This section wasn't part of Fuji's original design, but they thought it would add value to the profile page. Fuji dragged this section from the example UI onto the designated area of their current design \protect\ccircle{11}. 

\begin{center}
\includegraphics[width=\linewidth]{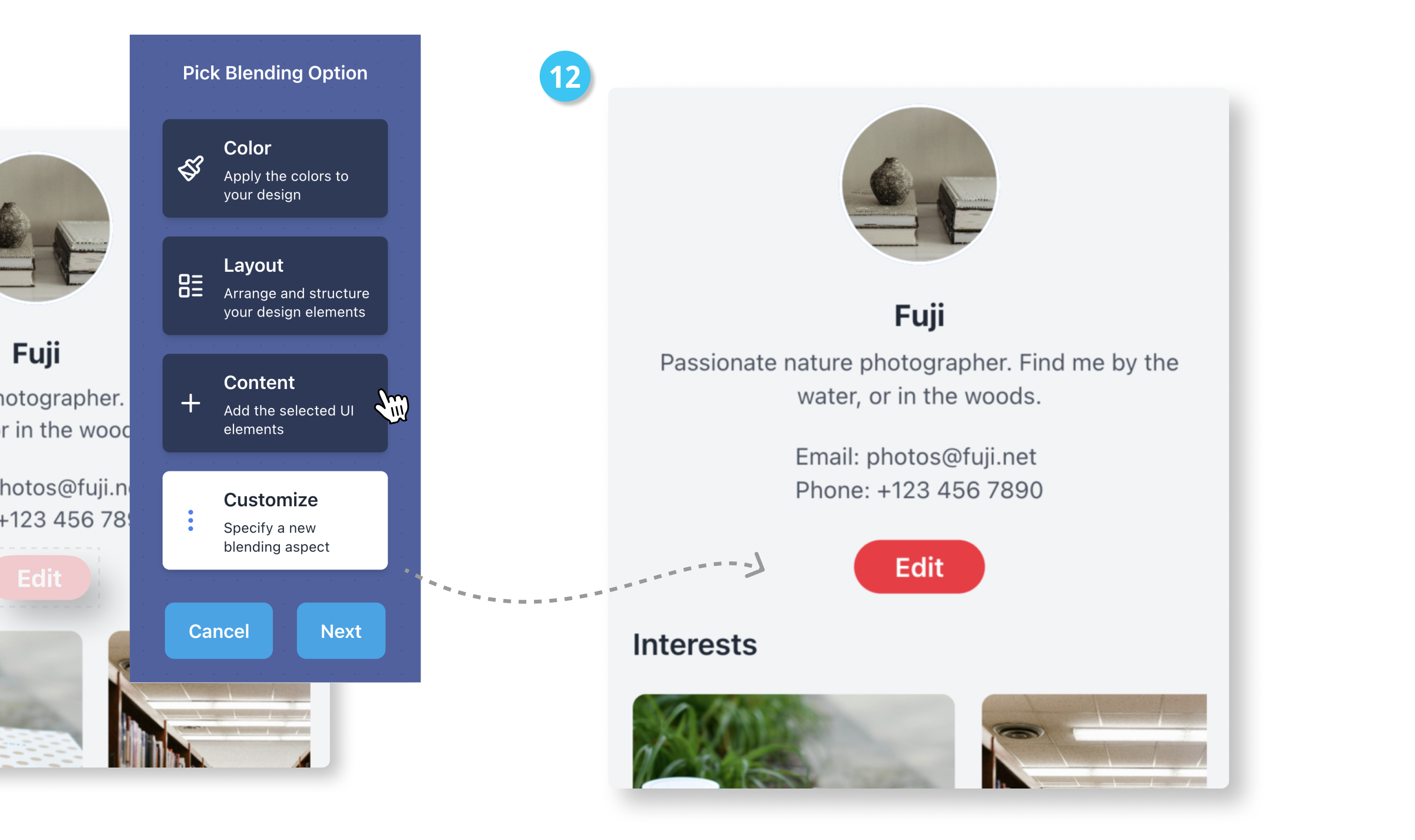}
\Description{This figure illustrates \sys{}'s blending process for integrating new UI elements. The image is divided into two parts, with a ``Pick Blending Option'' menu on the left and Fuji's updated profile on the right, labeled as 12. The blending menu shows options for ``Color'', ``Layout'', ``Content'', and ``Customize'', with the ``Content'' option highlighted. A dotted arrow connects this menu to Fuji's profile, which now includes a red ``Edit'' button beneath the contact information. The accompanying text explains that Fuji selected the ``color'', ``layout'', and ``content'' aspects to be blended from the example UI while maintaining the original design. \sys{} successfully implemented this request by adding the new ``Edit'' button to Fuji's UI at the specified location, demonstrating how \sys{} enables users to seamlessly incorporate new elements while preserving the overall design aesthetic of their existing interface.}
\end{center}

In the popup that appeared, Fuji specified the ``color'', ``layout'', and ``content'' aspects to be blended, preserving the original design in the example UI. \sys{} successfully added this new button into Fuji's UI at the specified location \protect\ccircle{12}.

Fuji is impressed with how quickly they were able to explore and iterate the UI using \sys{}. What would have taken hours of manual coding and design work , they were able to accomplish in a fraction of the time. The conceptual blending approach allowed Fuji to easily experiment with different design elements and see how they would look in the context of their UI. Moreover, Fuji appreciated that \sys{} didn't just provide a static design but generated UI code. They could now directly take the updated code back to their development environment. This seamless integration between design and code is particularly valuable for Fuji as a developer. Fuji realized that \sys{} can significantly speed up their UI development process in the future, allowing for rapid prototyping and iteration.


\section{Related Work}

In this section, we review prior work in example-driven UI prototyping, conceptual blending, and AI-enabled UI prototyping tools. These three lines of work characterize the main ideas behind \sys{} and build towards future tools that enable conceptual blending in UI/UX prototyping. 

\label{section:related-work}
\subsection{Using Examples in UI Prototyping}
Finding inspiration from examples is extremely common in UI ideation~\cite{herring2009getting, ritchie2011d, dow2010parallel, bunian2021vins, huang2019swire, chen2019gallery, lu2024flowy}. UI designers and developers use examples to understand best practices, gather inspirations, and refine design concepts. Online galleries like Mobbin and Dribbble\footnote{http://mobbin.com, http://dribbble.com} are useful in providing examples with different styles, components, and interactions. 

Prior research has introduced systems for retrieving design examples~\cite{huang2019swire, chen2019gallery, bunian2021vins, ritchie2011d, mozaffari2022ganspiration}. These systems use various methods, such as image recognition, generative networks, and metadata tagging to categorize and recommend design examples in a way that is easily searchable and accessible. Other prior works generate examples from constraints~\cite{swearngin2020scout} or layout diffusion models~\cite{cheng2023play, cheng2024colay}. \sys{} instead focuses on blending examples, but can potentially enhance these systems by providing an interface for UI developers for blending examples generated and presented by these tools.

More related to \sys{}'s core contributions are tools that visually transfer styles between UI designs~\cite{swearngin2018rewire, nguyen2015reverse, beltramelli2018pix2code,feng2022auto}. Uniform~\cite{nichols2006uniform} presents a framework for ensuring similarity between automatically generated cross-platform interfaces. Bricolage~\cite{kumar2011bricolage} helps web developers to transfer the content from one full webpage into the design of the other. Similarly, VST~\cite{warner2023interactive} supports interactive style transfer between vector graphics. These approaches present algorithms for computing semantic correspondances between design elements. Compared to \sys{}, these works have primarilly transferred the \textit{visual styles} between designs. With \sys{}, we expand the transfer to both visuals and content between UIs. We also enable \textit{interactive} transfer, where developers can iteratively specify sections and aspects of an example screen to transfer. While prior works~\cite{kumar2011bricolage, warner2023interactive} require code and SVG as input, \sys{} only requires screenshots from developers. This significantly expands the set of examples developers can use and makes \sys{} more practical in real workflows. 


\subsection{Conceptual Blending}

Conceptual blending is a prevalent theory in cognitive linguistics, establishing \textit{blending} as a cognitive mechanism for the construction of meaning in people's everyday life~\cite{fauconnier2003conceptual, camara2007creativity}. It involves identifying common elements between two concepts, selectively combining their features, and creating a new concept that inherits and transforms aspects of both inputs.\looseness=-1

Previous studies have demonstrated conceptual blending's utility in computational creativity support for various creative domains~\cite{camara2007creativity, zbikowski2018conceptual}. In music, conceptual blending has been applied to inspire music production by blending unique concepts drawn from music and language, two very distinct domains~\cite{zbikowski2018conceptual}. Conceptual blending has also been used as the basis for studying creativity and building computational implementations of creative acts~\cite{camara2007creativity}. Tools such as CreativeConnect have established blending as an effective operation in creative exploration for graphic design~\cite{choi2024creativeconnect}. \textcolor{orange}{Researchers have also designed direct manipulation interfaces for AI to support creative analogical reasoning~\cite{srinivasan2024improving}.}

Conceptual blending is intrinsic to the inspiration-seeking processes in UI prototyping~\cite{herring2009getting, lu2024flowy}. Current practices such as moodboarding also reflects conceptual blending's value for exploration and ideation~\cite{chen2019gallery, ritchie2011d}. In this work, we apply conceptual blending to UI development. We built our system \sys{} on the insight that the latest multi-modal AI models are capable of interpreting user intents, writing UI code~\cite{wu2024uicoder, si2024design2code}, and have the potential to operationalize conceptual blending because they encode certain characteristics of human ingenuity---for example, in metaphor reasoning~\cite{tong2024metaphorunderstandingchallengedataset, lakoff2014mapping} and conceptual blending of pop culture images~\cite{10.1145/3544548.3580948}. 

\subsection{AI Support for UI/UX Prototyping}
\label{sec:related-work-ai-support}
Many researchers have investigated the use of AI in UI design workflows~\cite{lu2022bridging, lu2024aib, knearem2023exploring, li2024user}. Emergent startups including \hyperlink{https://usegalileo.ai/}{Galileo}, \hyperlink{https://magicpatterns.com}{Magic Patterns}, and \hyperlink{https://v0.dev}{Vercel} have also demonstrated compelling uses of AI in UI design, yet their interactions are mostly restricted to chat interfaces and natural language instructions. 


For UI development, large multi-modal AI models have exhibited capabilities in code generation given UI screenshots~\cite{si2024design2code} or natural language descriptions~\cite{wu2024uicoder, lu2023ui, li2023using}. While techniques for code generation from natural language have advanced rapidly in recent years~\cite{yin2017syntactic, le2020deep, li2023starcoder}, generation based on provided UI design screenshots remains limited to simplistic outputs~\cite{si2024design2code}. 

Our system \sys{} expands existing work in two main directions. First, \sys{} does not adopt a chat-based interface, but primarily uses direct manipulation for users to specify their blending intents (i.e. which sections and aspects to blend in). Moreover, instead of code generation based on only screenshots or natural language descriptions, \sys{}'s usage scenario provides two main input sources: developers existing code (i.e. work-in-progress), and example UI screenshots to be blended into the code. This provides a more well-scoped, multimodal code editing task to fully utilize AI models' code generation capabilities.

\section{The \sys{} System}

To evaluate how conceptual blending can help developers prototype user interfaces, we developed \sys{} as a web-based system that supports interactive conceptual blending.

\label{section:system-code-representation-general}



\subsection{System Design Goals}

%
\sys{} embodies three design goals supported by findings from a literature review of three key areas: UI/UX design processes and tools~\cite{kumar2011bricolage, wu2023screen, warner2023interactive, lu2022bridging, knearem2023exploring, lu2024flowy, kim2022stylette, peng2024designprompt}, AI for creativity support~\cite{choi2024creativeconnect, chung2023promptpaint, angert2023spellburst, chen2024zero}, and Human-AI Interaction~\cite{horvitz1999principles, masson2024directgpt, shneiderman1997direct, subramonyam2023bridging, amershi2019guidelines}. We formulate the design goals as follows:

\begin{enumerate}

    %
    \item[\textbf{DG1}] \textbf{Support flexible and familiar input formats.} Because users employ representations at varying levels of fidelity depending on design focus~\cite{buxton2010sketching, norman2013design, huang2019swire} (e.g., sketches, wireframes, or high-fidelity prototypes), users should be given an input mechanism flexible enough to express intent across those representations. We identify UI examples in the form of screenshot images and hand-drawn sketches as natural mediums that agrees with common developer practices~\cite{herring2009getting}.
    

    \item[\textbf{DG2}] \textbf{Preserve developer agency in creative interaction.} Maintaining designer agency is crucial in AI-assisted design tools~\cite{lu2022bridging, knearem2023exploring, li2024user}. By offering granular control over the blending process, we ensure that the tool remains predictable and aligned with the user's vision, addressing key concerns raised in previous studies on AI adoption in design workflows~\cite{lu2022bridging, knearem2023exploring}. Studies have also shown that users prefer visual interactions over purely textual prompts when working with AI tools~\cite{subramonyam2023bridging, masson2024directgpt, lu2024flowy}.
    
    \item[\textbf{DG3}] \textbf{Promote exploration, reflection, and iteration.} Creative design processes are inherently iterative and reflective~\cite{buxton2010sketching, norman2013design}, so conceptual blending should support reflection by allowing the user to visually compare different blended results and to critically evaluate choices before making further changes. This iterative approach aligns with established design practices, such as parallel prototyping~\cite{choi2024creativeconnect, dow2010parallel, peng2024designprompt}, that help users refine ideas to arrive at more innovative solutions.
\end{enumerate} 

\subsection{Interaction Affordances}

\sys{} realizes the above design goals through three primary interaction affordances, which we now describe.

\subsubsection{Conceptual blending control for implicit intent disambiguation}

A primary challenge in \sys{}'s interaction design is the disambiguation of user intent, as the space of valid results for a conceptual blending operation is potentially large. For example, when a user wants to blend an example with their source UI, they might intend to blend the color, layout, content, or a combination of these aspects. Additionally, the user might only wish to blend from specific sections of the example, rather than the entire screen. 

To address this challenge, \sys{} provides interactions that implicitly disambiguate user intent. Concretely, \sys{} offers two primary blending modes, each respectively providing different levels of user control:

\begin{enumerate}
    \item \textbf{Global blending through node connection}: Users connect \exampleNode{}s to \sourceNode{}s, triggering a global, screen-level blending operation. This mode is suitable for more exploratory phases of design and exploits the language model's implicit world knowledge from pre-training~\cite{achiam2023gpt}.
    \item \textbf{Localized blending using drag-and-drop}: For more refined control, users can drag subsections of example UIs onto specific areas of the source UI. As the drag operation hovers over parts of the source UI, a highlight animation occurs which indicates affected areas and appropriately communicates expectations of the operation's outcome to the user~\cite{norman2013design}. Finally, upon dropping, \sys{} presents the user a dialog to specify which aspect(s) to blend.

\end{enumerate}

While global-level conceptual blending is suitable for exploratory phases of design, it can yield a low sense of control for users with specific goals in mind. \sys{}'s drag-and-drop interface for targeted blending operations allows users to precisely specify which elements they wish to blend. 

\subsubsection{Node-Based Interface and Infinite Canvas}
\sys{} employs a node-based system interface with two primary node types: \exampleNode{}s (\protect\ccircle{A} in Figure \ref{fig:misty-teaser}) showing example images and \sourceNode{}s rendering the source UI code (\protect\ccircle{B} in Figure \ref{fig:misty-teaser}). During a conceptual blending operation, a user blends an \exampleNode{} into a \sourceNode{}, creating a new \sourceNode{} with a rendering of the updated code. This node-based approach facilitates parallel prototyping~\cite{dow2010parallel} via exploration, visual comparison, and iteration of design alternatives. The infinite canvas, as also found in contemporary design tools such as Figma, Sketch, and Freeform, provides a familiar setup for users (DG1). Moreover, this architecture helps the user to trace the evolution of ideas, as nodes represent different stages or variations of a design. By visually representing the design process, the node-based interface inherently supports creative thinking and iterative refinement (DG3).\looseness=-1

We additionally embed a live code editor, expandable and collapsable to the right side of the screen, for editing the code underlying any \sourceNode{}. In this way, developers can apply manual tweaks as desired~(DG2). Note that code editors are only for \sourceNode{}s and \emph{not} \exampleNode{}s. Crucially, \sys{} does not require access to the underlying code for example images, different from previous research prototypes~\cite{kumar2011bricolage, warner2023interactive}. Users need only upload example \emph{images}, which may be more easily obtained than code.

\subsubsection{Dynamic UI for Semantic Refinement}
After a blending operation, \sys{} presents a summary of its changes organized into \textit{``semantic diffs''}---groups of semantic changes to the source UI describing the outcome of a conceptual blending operation. \textit{Semantic diffs} extend the idea of program diffs~\cite{hunt1976algorithm} by organizing changes based on their semantic impact on the UI, not on their character-level changes to the underlying code.
%

As a motivating example, consider the impact of converting a UI to dark mode: while conceptually a simple change, its realization might require multiple distinct edits to various CSS styles corresponding to different UI components.
\sys{} summarizes the combination of these edits into a dynamic widget whose content is generated on-the-fly based on the meaning of the change. Crucially, the generated widget affords both coarse-grained control (enabling or disabling dark mode), or fine-grained control (varying individual colors). This approach provides the user with a high-level, conceptual understanding of the blending outcome, while also preserving user control at different levels of granularity.
%

This implementation of semantic diffing, inspired by recent work in dynamic UI generation~\cite{tseng2024keyframer, cheng2024biscuit, vaithilingam2024dynavis}, ensures that users maintain agency throughout the blending process (DG2). It allows for rapid exploration of design alternatives while preserving the ability to make precise adjustments, thus supporting both divergent and convergent thinking in the design process (DG3).

\paragraph{Combined affordances.} The above affordances, when combined, create a spectrum of user agency ranging from full agent automation (global blending) to full direct manipulation (dynamic UI widgets and manual code editing)~\cite{shneiderman1997direct} in which the user can flexibly switch between levels of control (DG2). Global blending allows broad exploration in early stages, while granular controls (localized blending and semantic diff) allow refinement of specific aspects of the design at a later stage (DG3).
%
%
By offering this range of interaction options, \sys{} maintains the users' sense of agency and control throughout the process~\cite{satyanarayan2024intelligence} (DG2).


\subsection{Implementation Details} 

\sys{} is implemented in React, React Flow, Tailwind CSS, and React Live, with OpenAI's GPT-4o as the underlying large multimodal model servicing prompt requests. We employ few-shot prompting~\cite{brown2020language} with chain-of-thought reasoning~\cite{wei2022chain} to improve response quality (see \cref{appendix-prompts} for full details on the prompts used).
To detect and repair malformed React syntax, \sys{} employs heuristic rules implemented using regular expressions. However, in cases where these heuristics are insufficient, we note that users may opt to either re-generate the code or use the code editor to apply manual repairs.

To enable conceptual blending in \sys{}, we used GPT-4o~\cite{achiam2023gpt}. GPT-4o's capabilities in code generation, multimodal input processing, and broad world knowledge make it suitable for context-aware conceptual blending across various UI design domains. 
We asked the AI model to generate \textit{semantic diff} summaries together with each blending outputs. Afterwards, when the user toggle on/off each \textit{semantic diff} options, \sys{} triggers another API call to GPT-4o to update the code.

\subsection{System Limitations}

Our implementation of \sys{} focuses on UI content and visuals. 
These aspects are foundational elements of UIs; in the future, we plan to build on these static aspects and expand conceptual blending to UI interactivity. Conceptual blending for interactive UX prototypes presents its own unique challenges and opportunities, as it involves dynamic states and transitions between multiple screens. 

While GPT-4o and other large pre-trained multimodal models are capable and low-cost backbones for prototyping, it is likely that a custom model fine-tuned specifically for conceptual blending would offer more precise results that better follow user expectations. Recent technical advancements have shown great promise in creating a customized model in this direction~\cite{wu2024uicoder, shin2021multi}. However, developing such a model is beyond the scope of our current study. We identify this as a direction for future work (see Section \ref{sec:discussion}).

\section{Exploratory First-Use Study}
\label{sec:user-study}
To evaluate \sys{}, we conducted an exploratory, first-use user study with 14 experienced frontend engineers. Our study answers three primary research questions:

\begin{enumerate}
    \item[\textbf{RQ1}] How do developers utilize the different features in \sys{}?
    \item[\textbf{RQ2}] To what extent do \sys{}’s outputs satisfy developers’ expectations of conceptual blending?
    \item[\textbf{RQ3}] How does \sys{} and conceptual blending support developers' workflow?
\end{enumerate}


\subsection{Participants}
\label{sec:user-study-participants}
For our study, we recruited 14 frontend developers who knew React and Tailwind CSS. Participants worked at a large US technology company and were recruited via internal communication channel advertisements and word of mouth. Our participants varied in frontend development expertise, ranging from two to fifteen years of experience and novice to advanced knowledge of React and Tailwind CSS. \cref{tab:participants} shows a detailed breakdown of participant demographics. We carried out two study sessions in person and fourteen online, based on the participants' preferences. Study sessions lasted 60 minutes, and we compensated each participant with a \$12 digital gift card. 

\begin{table}[h]
\caption{The participant demographics of our user studies. \textmd{All participants are frontend developers with at least basic knowledge of React and Tailwind CSS, and experience in creating and improving the design of UIs.}}
\label{tab:participants}
\centering
\begin{tabularx}{\columnwidth}{>{\raggedright\arraybackslash}p{0.5cm} >{\raggedright\arraybackslash}p{0.8cm} >{\raggedright\arraybackslash}p{3cm} >{\raggedright\arraybackslash}X}
\toprule
\textbf{ID} & \textbf{YOE} & \textbf{Main Frontend Tools} & \textbf{React \& Tailwind} \\
\midrule
P1  & 8   & React                    & Advanced     \\
P2  & 7   & React                    & Novice       \\
P3  & 14  & React, TailwindCSS        & Advanced     \\
P4  & 12  & ReactJS, SvelteJS         & Proficient   \\
P5  & 2   & React                    & Intermediate \\
P6  & 5   & React, TypeScript         & Proficient   \\
P7  & 13  & React, TailwindCSS        & Advanced     \\
P8  & 4   & SwiftUI, React            & Advanced     \\
P9  & 11  & React, Vue               & Proficient   \\
P10 & 11  & React, NextJS            & Proficient   \\
P11 & 11  & React, TailwindCSS        & Advanced     \\
P12 & 15  & React, TailwindCSS        & Advanced     \\
P13 & 13  & React Ecosystem           & Advanced     \\
P14 & 5   & SwiftUI, React            & Advanced     \\
\bottomrule
\end{tabularx}
\end{table}

Since \sys{} uses frontend code to represent UIs, we recruited frontend engineers as our participants. Thus, they were able to fully use all features of \sys{}, especially the code editor, and understand the limitations of current AI models in generating frontend code~\cite{si2024design2code}. All participants had experience creating and improving UIs by themselves, rather than only implementing designs created by designer colleagues (see more discussion in \cref{section:system-code-representation-general}).

\subsection{Procedure}

We divided our study into the following three steps.
When participants performed the tasks, we asked them to think aloud to provide insights into their cognitive process\footnote{\cref{appendix-interview} and \ref{appendix-questionnaire} provide a detailed list of questions in our interview and usability questionnaire.}~\cite{ericsson1980verbal}.
We also piloted the study twice to refine the process and validate the tasks were realistic and engaging. \textcolor{orange}{These two pilot studies were not included in the above 14 participants.}



\subsubsection{Step 1 (10 min): Understanding current practices on using examples}

We began by exploring participants' current practices for finding and using design examples. We asked participants about their use of examples in their work, how they find them, and examples that significantly inspired their work. We also asked about their workflow to incorporate these examples into their projects.

\subsubsection{Step 2 (35 minutes): Use \sys{} to perform frontend development tasks through conceptual blending}

We then introduced participants to \sys{} and asked them to perform four frontend development tasks through conceptual blending:

\begin{enumerate}
    \item[\textbf{T1}] Blending a full example UI into the current source UI,\looseness=-1
    \item[\textbf{T2}] Adding the header section of an example UI to the source UI,
    \item[\textbf{T3}] Updating the source UI layout by blending in a hand-drawn screen sketch,
    \item[\textbf{T4}] Looking for examples online, and open-ended improvement of the source UI through conceptual blending, until satisfied.
\end{enumerate}

We designed these tasks to progress from \textit{constrained} to \textit{open-ended} exploration for the participants. The initial tasks familiarized the participants with \sys{}'s core functionalities, while the later tasks encouraged application of \sys{} to more subjective and open-ended design scenarios. The final open-ended task helped us understand how participants might naturally incorporate \sys{} into their existing workflows. This progression lets us naturally onboard our participants while assessing \sys{}'s usability, functionality, and potential impact across various design scenarios. 
Participants iterated on each task until they were satisfied with the results. We picked three representative UI layouts and components (i.e., book lists, information cards, phone apps) as the basis of these tasks. We asked the participants to perform the first three tasks using one example, but we asked them to pick a different example for the last task to have a clean slate for open-ended exploration.

After each blending operation, participants provided (1) a 5-scale value of how much the results met their expectations, and (2) qualitative feedback on ways to improve the results. These results helped us further understand developers' expectations and current multimodal AI models' limitations in UI conceptual blending.

\subsubsection{Step 3 (15 minutes): Usability questionnaire and follow-up interview}

After completing the tasks, participants completed a questionnaire on their experience using \sys{}. We adapted the questionnaire from the standard System Usability Scale~\cite{brooke1996sus} and included 5-point Likert scale questions evaluating \sys{} on its usability, user control and agency, output quality, value for design exploration, and real-world adoption potential. After completing the questionnaire, we conducted a semi-structured interview with participants. The interview questions covered scenarios where participants would or would not use \sys{}, desired improvements or additional features, concerns about using \sys{}, and thoughts on how tools like \sys{} might influence their work.




\section{User Study Results}

In this section, we provide the visualizations of participants' conceptual blending operations (\cref{fig:activity-trace}) and their responses to the usability questionnaires (\cref{fig:questionnaire}). We also conducted a qualitative analysis of the user study sessions' transcripts. These results together helped to answer our research questions regarding the value of \sys{} in UI development workflows.

\subsection{Quantitative Results}

\begin{figure}
    \centering
    \includegraphics[width=0.9\linewidth]{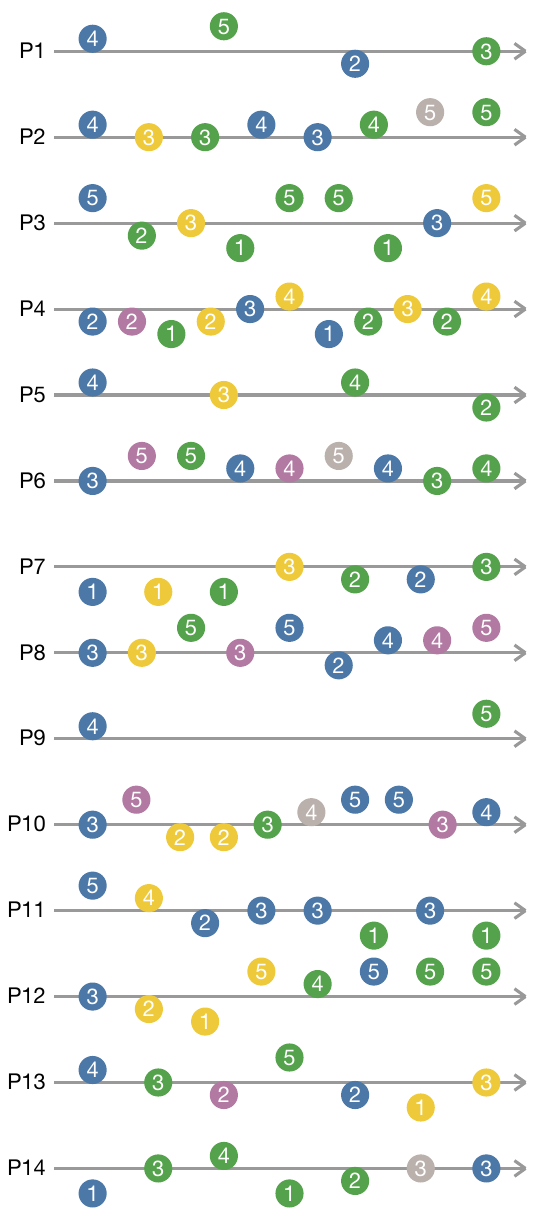}
    \caption{Event trace of all blending operations and user evaluation scores during the user study sessions. \textmd{Each circle represents a blending operation, laid out from left to right based on the time it occurred during the study. The number inside is the participant's evaluation of how well the blending output matched expectations, with 1 as not satisfied and 5 very satisfied. Based on these evaluation scores, we also varied the circles' vertical positions. The color of each circle represents the type of operation in \sys{}: \protect\globalcircle{} Global blend, \protect\dragcircle{} Drag and drop, \protect\semanticcircle{} Semantic diff, \protect\regeneratecircle{} Regenerate, \protect\editcircle{} Code edit. We normalize the space between operations.\looseness=-1}}
    \label{fig:activity-trace}
    \Description{This figure shows activity logs for 14 participants (P1 to P14) in a user study. Each horizontal line represents a participant's session, with colored circles along the line indicating different operations performed over time. The circles contain numbers from 1 to 5, representing the participant's satisfaction score for each operation (1 being least satisfied, 5 being most satisfied). The vertical position of each circle also corresponds to this score, with higher scores placed higher on the line. Different colors of circles represent different types of operations. Blue circles are most common, followed by green circles as the second most frequent. Yellow circles appear occasionally, while purple and gray circles are less common. The number and spacing of circles vary between participants, showing different patterns of activity and satisfaction across sessions. Some lines have many circles clustered together, while others have fewer, more spread out circles. The image provides a visual representation of user interactions and satisfaction levels throughout the study, allowing for comparison between participants and identification of patterns in operation types and satisfaction scores.}
\end{figure}

To answer RQ1 and RQ2---how developers utilize \sys{}'s different features and to what extent \sys{}’s outputs satisfy developers’ expectations---we visualized the blending operations conducted by participants in \cref{fig:activity-trace}. The average time each participant spent interacting with \sys{} was 36.3 minutes (median 36, sd 7.0). On average, each participant conducted 7.4 blending operations during the session (median 8, sd 2.5). Some participants conducted fewer blending operations (e.g. P1, P5, P9), because they provided extensive feedback and rationale. For the total of 103 blending operations, participants used Global blending (36/103) and Drag-and-drop (35/103) the most, constituting 34.95\% and 33.98\% of all operations, respectively. Users also used Regenerate (19/103, 18.5\%), Semantic diff (9/103, 8.74\%), and Code editing (4/103, 3.88\%) during the study.\looseness=-1

\begin{figure*}
    \centering
    \includegraphics[width=0.8\linewidth]{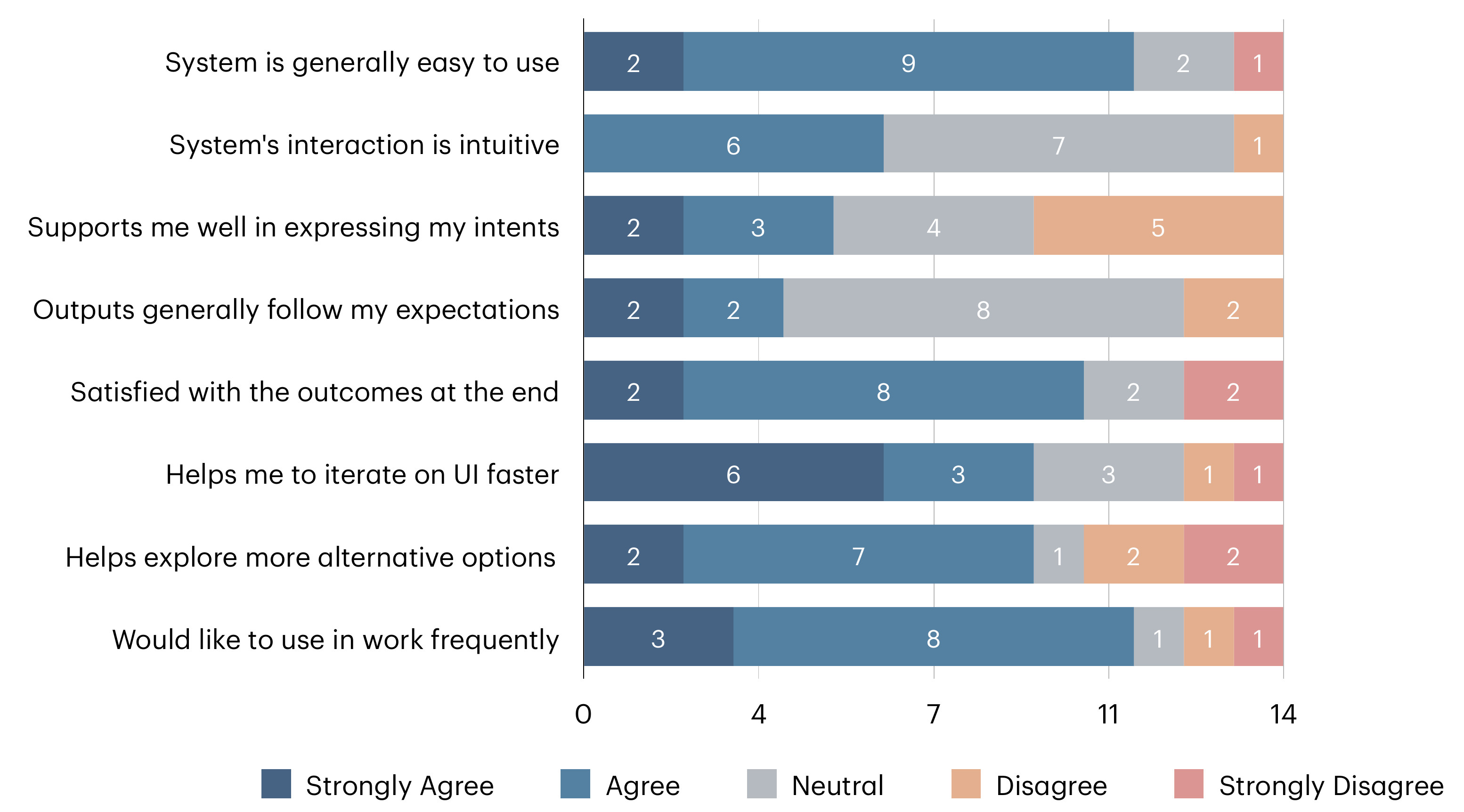}
    \caption{The usability questionnaire results from our user study. \textmd{User satisfaction survey results for \sys{}: most participants found \sys{} easy to use and would like to use it in their work regularly. \sys{} is generally well-received, with potential for targeted improvements in user intent interpretation and interface refinement.}}
    \label{fig:questionnaire}
    \Description{This image displays a horizontal bar chart showing the results of a usability survey for a system. The chart presents responses to eight statements about the system, with answer options ranging from ``Strongly Agree'' to ``Strongly Disagree''. The survey covers various aspects of the system's usability. Regarding ease of use, 11 out of 14 respondents agree or strongly agree that the system is generally easy to use, with only one strong disagreement. The system's interaction is considered intuitive by 6 respondents, while 7 remain neutral. In terms of supporting user intents, opinions are mixed. Only 5 respondents agree or strongly agree that the system supports them well in expressing their intents, with an equal number disagreeing. Similarly, 4 respondents feel the outputs generally follow their expectations, while 8 remain neutral. Satisfaction with outcomes is relatively high, with 10 respondents agreeing or strongly agreeing, and only 2 strongly disagreeing. The system appears to help with UI iteration, as 9 respondents agree or strongly agree that it helps them iterate on UI faster. The system also seems beneficial for exploring alternatives, with 9 respondents agreeing or strongly agreeing that it helps explore more alternative options. Finally, there's a strong inclination towards frequent use in work, with 11 respondents agreeing or strongly agreeing they would like to use the system frequently at work. Overall, the chart indicates generally positive reception of the system, particularly in terms of ease of use and desire for frequent use. However, there are areas with more divided opinions, notably in how well the system supports expressing user intents, suggesting potential areas for improvement.}
\end{figure*}

After each blending operation, participants also evaluated how satisfied they were with each output on a 5-point Likert scale from ``Very unsatisfied (1)'' to ``Very satisfied (5)''. The average satisfaction across all participants was 3.2 (median 3, sd 1.3). Out of all 103 activities, 31 received a satisfaction score below 3 (Unsatisfied or Very unsatisfied), 44 received a score above 3 (Satisfied or Very satisfied), with the remaining 28 rated 3 (Neutral). We noticed that some participants primarily expressed negative valence towards the outputs, for example, P7 and P14. These were mostly due to their high expectations of the outputs' visual details, or the random nature of \sys{}'s underlying model's behavior. Meanwhile, certain participants (e.g. P2, P6) showed generally positive valence towards the outputs. We observed that users' responses to the usability questionnaire (\cref{fig:questionnaire}) were influenced by their blending output quality in \sys{}.

Nine out of the fourteen participants rated some blending outputs as ``Very satisfied'', where \sys{} followed their expectations and captured the essence of examples in the updated code. 
Meanwhile, some blending outputs did not meet the users' expectations, mostly because they either did not reflect the users' expected changes or created UIs with non-ideal layout or alignment. 

Regarding the usability questionnaire responses (\cref{fig:questionnaire}), most participants find \sys{} generally easy to use and would like to use it frequently in their work. However, opinions were mixed on how well it supported expressing user intents, with only 5 out of the 14 participants (35\%) agreeing or strongly agreeing. Follow-up questions revealed that participants' main concerns for this question were centered around usability issues in the prototype, such as cumbersome interaction details in drag-and-drop and imperfect feature discoverability. Moreover, participants explained that the strong disagreement responses came from receiving \sys{} outputs that failed to follow their instructions or didn't meet their expectations for visual details (e.g., alignment, element sizing). These undesired outputs also reflected as low satisfaction scores in \cref{fig:activity-trace}.

\textcolor{orange}{For most participants who rated low on ``\sys{} supports me well in expressing my intents'', they requested features to continuously improve \sys{}’s output after conceptual blending. For example, after receiving an output, participants wanted to ask the AI model to further iterate on output details. These iterations do not involve reference examples, and are therefore not directly related to our focus on conceptual blending. }

\textcolor{orange}{When  specifically asked about \sys{}’s features for intent expressions, such as cropping and drag and drop, participants did not raise any issues. They felt these features addressed their needs in blending operations, as we will discuss later in \cref{theme-specify-intent}.}


\subsection{Qualitative Themes}
To answer RQ3---understanding how \sys{}'s interactive conceptual blending workflow supports developers---we conducted a qualitative analysis of our user study by transcribing the interviews, segmenting the interview content into quotations, and organizing these quotations into meaningful themes.  Our analysis resulted in six themes, and we present a summary of these themes in \cref{tab:themes}.

\begin{table*}
  \renewcommand{\arraystretch}{0.8}
  \centering
   \caption{Summary of qualitative participant feedback, organized by themes.\label{tab:themes}}
  \begin{tabular}{@{} >{\raggedright}p{0.15\textwidth} >{\raggedright}p{0.27\textwidth} >{\raggedright}p{0.34\textwidth} p{0.1\textwidth} @{}}
    \toprule
    \textbf{Theme}    & \textbf{Description} & \textbf{Representative Examples} & \textbf{Participants}                                            \\
    \midrule
    \emph{Kickstarting nascent creative explorations} (\cref{theme-kickstart-exploration}) & Participants found \sys{} valuable to accelerating early stages of UI exploration.                                         & 
    ``I would want to use this to do the grunt work of setting up something.'' \newline
    ``I would definitely use \sys{} for giving me a head start which I can later improvise on.'' \newline 
    ``It saves a good amount of time and effort from writing the code from scratch through any starting templates.''                                                           & P1--P4, P6, P10, P11, P14                                                 \\
    \emph{Allowing flexible intent specification} (\cref{theme-specify-intent}) & Participants appreciated the flexible intent specification mechanisms of \sys{} for various usage scenarios. &
    ``It’s very useful when I can't find any relevant templates to blend, I can just draw it as a sketch and use \sys{} to generate the UI.'' \newline
    ``The drag-and-drop is really good, it's basically an alternative to manually fixing it.'' \newline
    ``Being able to come in and toggle these on and off through dynamic widgets to make the AI output manageable like this is really cool.'' & P1--4, P6--12, P14              \\

\emph{Generating code as a blending substrate} (\cref{theme-codegen}) & Participants used the generated code in a variety of ways and noted their expectations of the generated code. & ``Designers can do the prototyping design phase and we can just take the code and use it.'' \newline ``This section of the code will be reused many times and will componentize.'' \newline ``I wouldn't necessarily expect it to make pixel perfect code, but I would something that I can then take from there and then bulletproof.''& P1, P3, P4, P12\\

   \emph{Inspiring through serendipity} (\cref{theme-serendipity})   &   Participants enjoyed unexpected blending outputs from \sys{}, which helps to provide inspirations they have not thought of. & 
    ``This one blend sparks joy because it is unexpected.'' \newline ``(\sys{} would be) useful to my work because it makes me think of things that I have not thought of.'' & P2, P5, P7, P8, P10, P14      \\
    
    \emph{Refining iteratively and visually} \newline (\cref{theme-refine-iteratively}) & After their initial explorations, participants desired additional capabilities to support further iterative refinement of their UIs. & ``I just want to tell it (\sys{}) to fix it without dragging and dropping” \newline ``A designer wants to be pixel perfect.'' \newline ``I'd love for the ability to add text to the prompt like I could when I was adding parts of a design to full styles.''                                    & P1, P2, P8--10 \\

    \emph{Blending boundaries between designers and developers} (\cref{theme-blend-boundary}) & Developers can see \sys{} as a tool for both themselves and designers, complementing each group with the other group’s skills, and blends the boundaries between these roles. & ``This blends the boundary between designer and developer.'' \newline ``It does a good job of covering for my lack of design skills.'' \newline ``(\sys{} is) raising the bar for developers of the app as far as the look and feel.'' \newline ``If designers use \sys{} to `generate some code’, I can use it immediately.''           & P1--4, P6, P12--14  \\
    
    \bottomrule
  \end{tabular}
\end{table*}
\renewcommand{\arraystretch}{1}


\subsubsection{Kickstarting nascent creative explorations}
\label{theme-kickstart-exploration}
All participants reported the use of examples in their UI development workflows. Developers would look for relevant UI examples online or find examples from component libraries. Then, they would often manually recreate these examples from scratch using the framework for their own projects (P3, P6, P8--10, P11, P14). The open source culture in web development---as well as browser inspection tools---allowed developers to easily ``borrow aspects of the design inspirations from others'' (P3) and build their UI from these starting ``patterns'' (P4).

\sys{} supports developers by scaffolding and accelerating exploratory UI prototyping in early stages of development workflows, especially ``being useful in very fast development'' (P3). This is triangulated by the usability questionnaire results, where most participants agreed that \sys{} helps to ``iterate on designs faster'' and ``explore more alternative design options''. Participants ``would want to use this (\sys{}) to do the grunt work of setting up something'' (P11). P4 estimated that \sys{} ``will add productivity boost to our work by 20\%--30\%'', since 30\%--40\% of their current work time is spent on creating CSS and static UI mockups. Participants prefer to take \sys{}'s output versus creating UI manually, as it can automate tasks that would take them 20--30 minutes to do otherwise (P10).

\subsubsection{Allowing flexible intent specification}
\label{theme-specify-intent}
Participants appreciated \sys{}'s diverse input mechanisms for specifying design intent, which supported various stages of their development process. The ability to blend in full screenshots, specific aspects of example UIs, and hand-drawn sketches proved efficient for different phases of UI development. For early ideation, participants valued the option to start exploring with generally blending in full screens or hand-drawn sketches. Particularly, hand-drawn sketches are widely used in roughly laying out design concepts and communicating design ideas. The ability to blend in sketches fits well with developers' current practices (P3, P6, P11, P12), especially ``in the case (when) I have an idea in mind, for which I cannot find any relevant template online'' (P6). P3 noted ``a lot of my time is just translating a sketch... this (\sys{}) can save me time on making the code or the visual look, so I don't have to fight with CSS and layout''. As designs progressed, participants found value in having more specific input options. P3 noted, for example, that drag-and-drop is very good for updating existing products, instead of starting brand new ones.

\subsubsection{Generating code as a blending substrate}
\label{theme-codegen}

Participants proposed utilizing the generated code in different ways, including as another avenue for refining the UI blending results (\cref{theme-refine-iteratively}) to ``easily fixing little things`` (P2) or adjust other ``minute aspects like font or spacing that's quicker to just do myself'' (P14). For some participants, authoring user interface code for presentation is tricky, and \sys{} can help developers avoid ``having to fight with CSS'' (P3) and the tool ``would help with a lot of the presentation boilerplate'' (P1).

Other participants desired to ``take the code and just use it`` (P4) or to uplift an ``existing legacy app interface to generate compatible but modern React and Tailwind'' (P3). P11 elaborated on their expectations around this code generation hand-off: ``With these kinds of systems, I wouldn't necessarily expect it to make pixel perfect code, but I would want the code that it resulted in to be something that I can then take from there and then bulletproof myself.'' 

A gap in \sys{} is that some participants wanted the code generation to emit UI components that map to their own libraries and design systems (P1, P5, P6, P10, P12). For example, P12 suggested organizing the output code into components, to better fit into the Atomic Design principles~\cite{frost2016atomic}: ``If it (\sys{}) could build that as a component, or at least tell me that this thing is repetitive (reusable), that would solve a major problem.'' In some situations where \sys{} complements designers, ``developers might not have the freedom to output any design they like'' (P6). A future version of \sys{} could allow preloading a set of design components for the tool to draw from or take advantage of custom data schemas---that is, ``being able to fill in the UI with my own list of data'' (P1)---when these are made available to the \sys{}.

\subsubsection{Inspiring through serendipity}
\label{theme-serendipity}
A benefit of \sys{} was its ability to generate surprising blending outputs that inspire developers. Six participants (P2, P5, P7, P8, P10, P14) reported enjoying serendipitous outputs that sparked new ideas. During the study, P2 received a particularly unexpected yet inspiring blend ``that sparks joy''. This sentiment was echoed by P5, who found \sys{}'s surprise blends ``to the most useful to my work, making me think of things that I have not thought of.'' These experiences suggest that \sys{} stimulates developers' creativity, pointing out new design directions outside of established patterns.

However, serendipity can also be a double-edged sword. Although some of the blends initially have an ``amazing wow factor'' (P10), when carefully scrutinizing the blend there are often small details that were initially overlooked, such as the ``separation between items and not apply the button design style to one of the buttons'' (P10). The randomness, which ``can be good in exploration'' (P7), is ``also bad when you already know where to go'' (P7) or ``have a specific idea in mind'' (P3). For such scenarios, having a ``more deterministic way of working with the tool would be preferred'' (P3). This experience made P3 ``impressed by some things and yet underwhelmed by others.''

To address these trade-offs, a future improvement could give a knob to the developer that allows them to indicate their preference for serendipity. Earlier in the design process they may set this preference higher, and lower it later in the design process when they know more precisely what design they want.

\subsubsection{Refining iteratively and visually}
\label{theme-refine-iteratively}
While \sys{} helps in initial explorations, participants needed more support for iterative visual refinement after the initial exploratory phases (P1, P2, P8--10). Participants often found themselves wanting to make precise adjustments directly over the output, without drag-and-drop of another example. Participants wanted the ability to ``add text prompt to the output, like I could when I was adding parts of a design to full styles'' (P8). This comes from a demand for pixel-perfect UI implementations in professional contexts, which current models are not always able to produce~\cite{si2024design2code, wu2024uicoder}. From many UI developers, they are expected to produce pixel perfect implementations, especially by designers (P3). As a result, when seeing issues in the blending outputs' visual details (e.g. element sizing, alignment, and color values), developers naturally wanted to fix them with direct prompt iterations over the visually rendered UI. This is currently only supported by direct editing in our code editor. However, developers also recognized that they are fine with manually tweaking these details. While the blending outputs were not pixel perfect, developers generally thought they were ``good enough output for me to build upon'' (P4), then, ``the minute interaction of copying over font or changing some spacing, that's something I can do way quicker by myself'' (P14). Yet, developers can still benefit from direct iterations on the visual outputs in \sys{}, without necessarily involving more examples.

\subsubsection{Blending boundaries between designers and developers}
\label{theme-blend-boundary}
P2, P4, and P13 considered both designers and developers as potential user groups that can benefit from using \sys{}. The system showed potential in blending designer and developer activities in UI prototyping, shifting collaboration and handoff points in the full prototyping workflow. This is noted by eight participants in our study. P1, P4, and P14 mentioned the shifting handoff points between designers and developers, where traditionally they only receive design artifacts that contain no code implementation. If designers use \sys{} to generate some code, developers can directly take the code and immediately use it in their implementation (P1). \sys{} can also help developers to incorporate stakeholder feedback faster, enhancing cross-function collaboration (P12) and reducing the friction with designers (P3, P13).\looseness=-1

At a higher level, \sys{} also complements designers and developers with each others' skills. P2 noted that \sys{} ``does not replace me, but does a good job of covering for my lack of design skills.'' \sys{} helps to raise the bar for design for developers' UI implementations. Participants imagined designers using \sys{} may not have to rely on engineers and vice versa (P11). In the future, tools like \sys{} will continuously lower the boundary of entry for both UI design and development, changing the separated dynamics and goals for these two roles.

\section{Discussion}
\label{sec:discussion}

\paragraph{Implications for responsible AI in conceptual blending tools}

Artists and creators have expressed concern over content ownership as generative AI models advance~\cite{srinivasan2021role, flick2022ethics}. Conceptual blending, while prevalent in creative tasks, is essentially borrowing content and styles from existing examples that users do not necessarily own. Conceptual blending tools have to be responsibly deployed with careful ethical considerations around ownership, permissions, and their use in blending derivatives.

UI/UX design and development pose unique challenges for these ethical considerations in conceptual blending. Since software users do not adapt quickly to new interface and interaction patterns, designers often need to incorporate established patterns that users are already familiar with~\cite{norman2013design}. As a result, usability---instead of novelty or creativity---is often prioritized in UI/UX prototyping. \textcolor{orange}{For this reason, the boundary of unethical appropriation in UI/UX design is often ambiguous. Moreover, designers and developers open-source component libraries for others to reuse and re-purpose.} 

\sys{} currently does not have safety guardrails around what inputs the system can accept and what outputs the system can generate. As a result, \sys{} can generate inappropriate user interfaces if such blends are requested by the developer. There are continuing conversations on how to balance developer freedom around flexible intent specification and when (and whether) to deny the generation of certain user interfaces. 

These are only some challenges that the generative UI community is grappling with that would need to be resolved in this evolving landscape of Generative AI technology. \textcolor{orange}{It would be valuable to prioritize responsible AI and copyright ethics to improve the design of \sys{}. Insights from this design approach, and its influence on user interaction with the tool would contribute to HCI and Responsible AI.}


\paragraph{Augmenting refinement with bespoke widgets}

In \sys{}, we demonstrated the feasibility and utility of ``semantic diff'' using a simple toggle widget \cref{fig:misty-teaser}. However, there exists an expressive yet underexplored design space of bespoke widgets that make semantically meaningful adjustments beyond direct attribute edits. For example, a \emph{layout density slider} can tweak screen element layouts from ``sparse'' to ``dense'', triggering simultaneous updates of CSS spacing attributes without breaking the harmony of the layout. A \emph{color scheme grid} can generate the entire page's color scheme---instead of single color pickers---along axes such as ``monochrome'' to ``multicolored'' and ``dark'' to ``light'' themes. We can also imagine a screen \emph{content verbosity slider} from ``concise'' to ``verbose'', together updating multiple pieces of text on screen.

To successfully materialize these bespoke widgets, \sys{} would need to incorporate additional context, such as system states and user intentions. Previous work has also discussed the potential value of serializing and saving these bespoke widgets for reuse~\cite{cheng2024biscuit}.

Nevertheless, introducing these bespoke widgets to future tools similar to \sys{} can support more semantic, iterative refinement that improves developer productivity.

\paragraph{Expanding conceptual blending in \sys{}}

Our implementation of \sys{} is based on simple UI screens with limited interactivity. During our user study, participants expressed interest in expanding \sys{} to conceptual blending of interactive prototypes. Interactive UI prototypes contain many dimensions that are relevant for conceptual blending: user interactions, animations, and screen states. Tools like CoCapture~\cite{chen2021cocapture} have made progress towards blending interactivity, and Keyframer~\cite{tseng2024keyframer} has investigated the use of Generative AI for authoring animations. However, blending interactivity continues to pose design challenges. Blending static screenshots and sketches are relatively simple, yet it is not obvious how to specify sections and aspects---for example, of a video---for blending. Design principles from research in programming by demonstration can provide a useful starting point~\cite{cypher1993watch}.

We also plan to expand \sys{} to support more ``unusual blends'', such as blending of non-UI media types---pictures, videos, and even music---into UI screens. These conceptual blends might be able to borrow inspirations based on aesthetic or mood. For example, an audio clip containing somber tones could result in the generation of a user interface with dark and muted colors. Similarly, a photograph from a science fiction film might result in a user interface that mimics the affordances appropriate to that sci-fi genre. As other forms of art are rich in metaphor and figurative language, it would be amusing to consider how Generative AI can meaningfully interpret such unusual blends to construct surprising user interfaces.


\section{Limitations}

\subsection{User Study Limitations}

Our study participants were frontend developers recruited from a single US-based technology company. Other roles, such as UI/UX designers, may have different experiences with conceptual blending using \sys{} than developers. US-centric participants may also limit our understanding of how different culture design preferences might affect usage patterns and UI outputs. The first-use study format and controlled task environment may not always reflect how \sys{} would be used in real-world scenarios. A longitudinal study of \sys{} being used in the wild can be a valuable future work to answer these questions.

\subsection{System and Technical Limitations}

\sys{} works with static, single-screen mobile UI. It does not support the conceptual blending of interactive prototypes and UI animations. Understanding multi-screen, interactive UI prototypes with AI has been an active research area~\cite{wu2023never, lu2024flowy, zhang2024interaction}. UI motion design protocols such as Lottie Files~\cite{lottiefiles} also make conceptual blending of UI animations easier. We expect to build \sys{} on top of these work to extend its blending capabilities.

\sys{} only supports accessibility incidentally~\cite{wobbrock2011ability, AppleHIGAccessibility, stephanidis1998universal, swearngin2024towards}. \sys{} can incorporate accessibility design best practices~\cite{AppleHIGAccessibility}---for example, by including post-hoc checking mechanisms to ensure accessibility---such as the ones in DesignChecker~\cite{huh2024designchecker}. Adopting an ability-based design approach, \sys{} can also generate blending outputs based on user abilities, to serve diverse users' needs.

\sys{} also does not consider cross-cultural design~\cite{plocher2021cross, hofstede2009geert, bourges1998meaning, kellogg1993cross} in conceptual blending. Research has shown that users from different cultures have diverse preferences for AI capabilities, control, and connections~\cite{ge2024culture}. Cultural contexts also influence how AI models and users influence each other's decision making~\cite{lee2020understanding}. \sys{} could adopt more culturally-aware AI models in generating blending outputs~\cite{suzuki2023we, robinson2020trust}. Co-designing and testing \sys{} with users from different cultures can also help us see how their preferences and desired outputs may differ~\cite{heaton2002designing}.

\subsection{Responsible AI Considerations in UI Prototyping}

Practitioners and researchers have been incorporating Responsible AI and AI ethics into technical systems~\cite{amershi2019guidelines, wright2020comparative, shneiderman2020bridging, wang2023designing, varanasi2023currently, ali2023walking}. Systems and toolkits have been proposed in HCI research to support Responsible AI in design and development workflows~\cite{munechika2022visual, cabrera2019fairvis, wexler2019if, wang2022interpretability}.

The boundary of ``ethical'' and ``unethical copying'' in UI/UX is fluid. Much of today's UI design and development builds upon established design patterns to prioritize usability over creativity~\cite{talton2012learning, lu2024flowy}. Also, open-source communities promote re-using and re-purposing pre-built UI components~\cite{bach2010social, zilouchian2012consensus, apple_design_resources, shadcn_ui}. 

Before AI adoption in design tools, we already witnessed increased homogenization of website design~\cite{goree2021investigating}. AI-enabled prototyping workflows, like \sys{}, can amplify this trend~\cite{anderson2024homogenization}. Automated prototyping tools raise questions about authorship and intellectual property. They risk copyright infringement and can spread biased design choices. Guidelines and ethical oversight can ensure these tools support creative freedom and minimize harm.

In \sys{}, we evaluated the experience of conceptual blending to understand its impact on UI prototyping workflows. \sys{} does not currently have safeguards. Design opportunities exist for \sys{} to prioritize Responsible AI considerations. For example, researchers can visualize an AI model's conceptual blending outputs, to help users understand the output distributions. Combined with visual analytics~\cite{wexler2019if, munechika2022visual}, this could help users discover and understand AI model biases. Furthermore, in-situ support similar to \textsc{Farsight}~\cite{wang2024farsight} could increase users' Responsible AI awareness in prototyping and adopting blending outputs. Future research in this direction can lead to research insights for conceptual blending in UI prototyping.
\section{Conclusion}

We built \sys{}, a system that embodies the conceptual blending and enables designers and developers to blend concepts from UI examples into their own work. We operationalized conceptual blending within \sys{} through a multi-modal AI model---postulating that these models have some capability to interpret and apply conceptual blending. Although imperfect and occasionally uneven in its generation, these models nevertheless demonstrated a remarkable ability to apply blends and even produce surprising and serendipitous results. Participants in our exploratory first-use study were encouraged by \sys{}'s capabilities. Creativity is a kaleidoscope, and \sys{} demonstrates the potential for tools that blur the boundaries between developers and designers.

\bibliographystyle{ACM-Reference-Format}
\bibliography{references}

\clearpage
\appendix
\section{Appendix}
\subsection{Interview Questions For User Study}
\label{appendix-interview}

Here, we provide a detailed list of questions asked to participants during the user study. Please refer to Section \ref{sec:user-study} for a comprehensive description of the user study procedures.

\subsubsection{Step 1: Understanding Current Practices on Finding Inspirations}
\begin{enumerate}
    \item How do you use inspirations in your work? If so, how do you find and use them?
    \item Can you show me some examples that you learned a lot from or are very inspired by? 
    \item How did you incorporate these examples into your work
\end{enumerate}

\subsubsection{Step 3: Follow-Up Interview Questions}

Please note that given it is an semi-structured interview, we sometimes adjusted the questions based on the participant's questionnaire response and question answers.
\begin{enumerate}
    \item What are scenarios you would like to use \sys{}?
    \item How will it influence your current workflow?
    \item What are scenarios you would not like to use \sys{}? Why?
    \item What do you think about the ability to blend in examples of different fidelities?
    \item If you do not only blend in examples, what are some other ways you might use conceptual blending?
    \item What are some of the ways \sys{} can be improved? What other features do you desire?
    \item How to improve \sys{} further to make it more useful?
    \item Any concerns using \sys{}?
    \item \sys{}’s features support design iteration, but are directly implemented using frontend code. As a developer, how do you feel? Will this influence your work? Will this impact designer-developer handoff?
\end{enumerate}

\subsection{Usability Questionnaire For User Study}
\label{appendix-questionnaire}
Each question is accompanied by a 1-5 likert scale.

\begin{enumerate}
    \item The system is generally easy to use.
    \item The system’s interaction is intuitive.
    \item The system supports me well in expressing my intents.
    \item The system’s outputs generally follow my expectations.
    \item I am satisfied with the outcomes I get at the end.
    \item The system can help me to iterate on UI designs faster.
    \item The system helps me to explore more alternative design options than I usually would.
    \item I think that I would like to use a system like \sys{} in my work frequently.
\end{enumerate}

\subsection{Generative AI Prompts}
\label{appendix-prompts}

Here, we provide two main types of prompts we used to generate conceptual blending outputs in \sys{}.

\subsubsection{Global Blending}

Global blending generally involves the work-in-progress code and a reference image (separately uploaded).

\lstset{
    breaklines=true,                 
    basicstyle=\ttfamily\footnotesize,        
    breakatwhitespace=false,         
    showstringspaces=false,          
    frame=none,                      
    breakindent=0pt,                 
    language=,                       
    escapeinside={(*@}{@*)},         
}

\begin{lstlisting}

Here is my react and tailwind code: 
                
        ${work-in-progress code} 

        Help me create a new component page that blends the content of the above code with the visual style, layout, and appearance of the reference image. Change only the source code corresponding to this, and no other sections.

        Preserve the content in the UI of the code, follow the layout and visual style of the reference image, optionally add more content where the original code's content cannot fill in all fields in the reference image's layout. Do not use content from the reference image, just use its layout and visual style. 

        A few rules:

        - First, explain in concise language the design of the reference screenshot. Use it as a basis of your generation.
        - Make sure all text is legible on the background
        - Briefly summarize the differences between the reference image and the code, summarize them into a few categories of changes you want to make. Base your later generation of categorizedChanges based on these categories.
        - only use tailwind, react, and react icons. Follow the current code structure, do not include any import or export statements, just use a simple component definition () => {};
        - when adding icons, pick from this list of lucide react icons, do not use others: ${list of available lucide icon names}
        - there are a few stock photos for use under the folder /stock/, they are named after their orientation, like landscape0.jpg, landscape1.jpg, portrait0.jpg, etc. Do not use any other images. Do not use placeholder image paths.
        - Summarize the code changes in your response, use the format "categorizedChanges:" followed by a list of changes. Be very concise in your explanations. For example, "Color change: section titles, from green to purple"; "Layout change: adapted the layout for [add the feature description of the changed code piece]".


        Return result in the below format, make sure you use json:

        {
            "designExplanation": // explain the design of the screenshot image, focus on layout and color, be really concise, less than 30 words
            "differences": // briefly summarize the differences between the reference image and the code, focus on layout orientation, spacing, color theme, font, etc.
            updatedCode: \`() => {}\`   // return the whole component for the entire screen, with the updates;
            // a list of objects listing the changes made, use the tailwind classes to indicate the changes
            categorizedChanges: [
                {
                    category: "",   // summarize the category of the below changes, group changes together semantically, e.g. "Color: Changed light to dark theme", "Layout: Increased spacing between elements", "Visual details: Increased corner roundedness", "Image: Decreased image size", "Font: Changed font appearance", etc.
                    changes: [{
                        type: "color",
                        before": // the tailwind class before the change,
                        "after": // the tailwind class after the change
                    }]
                }
            ]
        }

        here is a good example of the changes field:
        categorizedChanges: [
            {
                category: "Color: Changed light to dark theme",
                changes: [{
                    type: "color",
                    before: "bg-black",
                    after: "bg-white"
                }, {
                    type: "color",
                    before: "text-white",
                    after: "text-gray-900"
                }, {
                    type: "border",
                    before: "", // you can use empty before field to indicate addition of new classes
                    after: "border-2 border-gray-300/90"
                }, ...] // add as many as appropriate,
            },
            {
                category: "Font: Changed font appearance",
                changes: [{
                    type: "color",
                    before: "bg-black",
                    after: "bg-white"
                }, {
                    type: "font",
                    before: "text-sm",
                    after: "text-lg"
                }, ...] // add as many as appropriate,
            },
            
            ]
\end{lstlisting}

\subsubsection{Drag-and-drop Blending}
The drag-and-drop blending prompt differ from the global blending prompt mainly in two ways. First, it provides the desired aspects users want to blend in (e.g. color, layout, content). Second, it adds the specific target code that was dropped on. 

\begin{lstlisting}
% [breaklines,fontsize=\footnotesize,breaksymbol={}]{text}

Here is my react and tailwind code: 
                
        ${work-in-progress code} 

        blend the ${user specified blending aspect} of the reference image into ${targetCodeDropped}. Change sections of the source code corresponding to this, as well as sections that are of similar layout or screen position to this. For example, don't just apply to one element in a list, but apply to all list elements with similar layouts. Sometimes the specific code piece does not correspond to parts of the source code, because it's rendered HTML based on the source React code. In that case, you need to identify the original code pieces from the source and modify them.

        A few rules:


        - First, explain in concise language the layout of the reference screenshot. Use it as a basis of your generation.
        - Make sure all text is legible on the background.
        - Briefly summarize the differences between the reference image and the code, summarize them into a few categories of changes you want to make. Pay attention to ${blendMode.filter(mode => mode !== "Customize").join(" ")}. Base your later generation of categorizedChanges based on these categories.
        - Never directly pulls content from the reference to update the source code. For blending color and layout, preserve all original content in the UI for source code, only change/add the original content when it's really necessary for following a layout. When you blend in the addition mode, generate content based on the context of the source code.
        - Do not use list and .map functions to represent lists. Just generate HTML elements for each of the list items.
        - only use tailwind, react, and react icons. Follow the current code structure, do not include any import or export statements, just use a simple component definition () => {};
        - when adding icons, pick from this list of lucide react icons, do not use others: ${list of avaialble lucide icons}

        - there are a few stock photos for use under the folder /stock/, they are named after their orientation, like landscape0.jpg, landscape1.jpg, portrait0.jpg, etc. Do not use any other images. Do not use placeholder image paths.
        - Summarize the code changes in your response, use the format "categorizedChanges:" followed by a list of changes. Be very concise in your explanations. For example, "Color change: section titles, from green to purple"; "Layout change: adapted the layout for [add the feature description of the changed code piece]".
        - When creating a bottom navigation bar, use "absolute bottom-0" instead of "fixed bottom-0".
        - Try to make colors and styles consistent and harmonious with the rest of the component.
        

        Return result in the below format, make sure you use json:

        {
            "designExplanation": // explain the design of the screenshot image, focus on layout and color, be really concise, less than 30 words
            "differences": // briefly summarize the differences between the reference image and the code, focus on layout orientation, spacing, color theme, font, etc.
            updatedCode: \`() => {}\`   // return the whole component for the entire screen, with the updates;
            // a list of objects listing the changes made, use the tailwind classes to indicate the changes
            categorizedChanges: [
                {
                    category: "",   // summarize the category of the below changes, group changes together semantically, e.g. "Color: Changed light to dark theme", "Layout: Increased spacing between elements", "Visual details: Increased corner roundedness", "Image: Decreased image size", "Font: Changed font appearance", etc.
                    changes: [{
                        type: "color",
                        before": // the tailwind class before the change,
                        "after": // the tailwind class after the change
                    }]
                }
            ]
        }

        here is a good example of the changes field:
        categorizedChanges: [
            {
                category: "Color: Changed light to dark theme",
                changes: [{
                    type: "color",
                    before: "bg-black",
                    after: "bg-white"
                }, {
                    type: "color",
                    before: "text-white",
                    after: "text-gray-900"
                }, {
                    type: "border",
                    before: "", // you can use empty before field to indicate addition of new classes
                    after: "border-2 border-gray-300/90"
                }, ...] // add as many as appropriate,
            },
            {
                category: "Font: Changed font appearance",
                changes: [{
                    type: "color",
                    before: "bg-black",
                    after: "bg-white"
                }, {
                    type: "font",
                    before: "text-sm",
                    after: "text-lg"
                }, ...] // add as many as appropriate,
            },
            
            ]
\end{lstlisting}

\end{document}